\documentclass[twocolumn]{aastex63}

\newcommand{\g}{log($g$)}

\received{August 15, 2024}
\revised{September 30, 2024}
\accepted{October 16, 2024}
\submitjournal{ApJ}

\shorttitle{Icarus revisited: An Accreted Stellar Stream in the Disk of the Milky Way}
\shortauthors{Re~Fiorentin et~al.}

\begin{document}

\title{Icarus revisited: An Ancient, Metal-poor Accreted Stellar Stream in the Disk of the Milky Way
}

\correspondingauthor{Paola Re Fiorentin}
\email{paola.refiorentin@inaf.it}

\author[0000-0002-4995-0475]{Paola Re Fiorentin}
\affil{INAF - Osservatorio Astrofisico di Torino, Strada Osservatorio 20, I-10025 Pino Torinese (TO), Italy}

\author[0000-0003-1732-2412]{Alessandro Spagna}
\affil{INAF - Osservatorio Astrofisico di Torino, Strada Osservatorio 20, I-10025 Pino Torinese (TO), Italy}

\author[0000-0003-0429-7748]{Mario G. Lattanzi}
\affil{INAF - Osservatorio Astrofisico di Torino, Strada Osservatorio 20, I-10025 Pino Torinese (TO), Italy}
\affil{Department of Physics, University of Torino, Via Giuria 1, I-10125 Torino, Italy}

\author[0000-0001-6291-6813]{Michele Cignoni}
\affil{Department of Physics, University of Pisa, Largo Pontecorvo 3, I-56127 Pisa, Italy}
\affil{INFN, Largo B. Pontecorvo 3, I-56127 Pisa, Italy}
\affil{INAF - Osservatorio di Astrofisica e Scienza dello Spazio, Via Gobetti 93/3, I-40129 Bologna, Italy}

\author[0000-0001-5412-869X]{Sara Vitali}
\affil{Istituto de Estudios Astrof\'isicos, Facultad de Ingenier\'ia y Ciencias, Universidad Diego Portales, Av. Ejército Libertador 441, Santiago, Chile}
\affil{Millenium Nucleus ERIS, Chile}

\begin{abstract}

The search for accreted satellites in the Galactic disk is a challenging task, to which {\it Gaia} plays a crucial role in synergy with ground-based spectroscopic surveys. In 2021, Re~Fiorentin et~al. discovered five~substructures with disk kinematics including Icarus. To gain more insight into the origin of Icarus as a~remnant of a~dwarf galaxy rather than a~signature of secular processes of disk formation, we complement astrometric {\it Gaia}~DR3 data with spectroscopy from APOGEE~DR17 and GALAH~DR3, and explore the chemo-dynamical distributions within 3~kpc of the Sun. We select 622~stars in the accreted/unevolved regions of [Mg/Mn]-[Al/Fe] and [Mg/Fe]-[Fe/H], where we identify 81 and 376~stars with $-2<{\rm[Fe/H]}<-0.7$ belonging to Icarus and Gaia-Sausage-Enceladus (GSE), respectively. The revised properties of Icarus are:  
$\langle V+V_{\rm LSR}\rangle\simeq171~\rm{km~s}^{-1}$, 
$\sigma_{V}\simeq37\rm{km~s}^{-1}$, 
$\langle e\rangle\simeq0.36$, 
$\langle{\rm[Fe/H]}\rangle\simeq-1.35$, 
$\langle{\rm[Mg/Fe]}\rangle\simeq+0.27$, 
$\langle{\rm[Al/Fe]}\rangle\simeq-0.13$, and 
$\langle{\rm[Mn/Fe]}\rangle\simeq-0.39$.
From the CMD of its members, Icarus appears older than 12~Gyr.
Such age and dynamical properties are reminiscent of the metal-weak thick disk. However, detailed chemical analysis in the diagnostic spaces [Ni/Fe]-[(C+N)/O], [Y/Eu]-[Fe/H], [Eu/Mg]-[Fe/H], [Ba/Y]-[Fe/H], and [Ba/Mg]-[Mg/H] evidences that Icarus and GSE occupy the accreted region, well separated from the bulk of {\it in situ} disk stars. Updated comparisons with N-body simulations confirm that Icarus' stars are consistent with the debris of a dwarf galaxy with a stellar mass of $\sim~10^9~M_\sun$ accreted onto a primordial disk on an initial prograde low-inclination orbit. 

\end{abstract}
 


\section{Introduction}\label{sec:1}
Understanding the formation processes of 
galaxies is one of the fundamental problems in astrophysics. 
In the standard $\Lambda$CDM model, 
galaxies grow by hierarchical merging, through the progressive assembly of matter. 
Simulations based on this paradigm confirm the build up of galaxies by accretion of smaller systems like dwarf galaxies, that interact through the action of gravity. 
As a result, the orbiting satellite galaxy is distorted and progressively disrupted by tidal forces exerted by the main system 
\citep[][]{Johnston1998, Bullock2005, Cooper2010, Fattahi2020}. 

The theory has gained substantial observational support thanks to the modern surveys at high redshift. 
In addition, our own Galaxy, the Milky Way (MW), constitutes an unique laboratory for studying the processes of formation and evolution of spiral galaxies in a local cosmology context \citep[e.g.,][]{Freeman}. 

In fact, extended tidal tails of bound dwarf galaxies and globular clusters 
have been detected in the Galactic outer halo \citep[e.g.][]{Malhan2018}, while in the solar neighborhood, where spatial overdensities are no longer coherent, debris from accreted satellites are observed as overdensities in velocity-space \citep[][]{Helmi1999, Smith2009, Klement2010, ReFiorentin2015, Naidu2020, Buder+2022}.

The MW inner halo is dominated by debris of Gaia-Sausage-Enceladus (GSE), a massive dwarf galaxy accreted $\sim 10~\rm{Gyr}$ ago \citep[][]{Belokurov2018, Helmi2018, DiMatteo2019, Gallart2019} 
and a metal-rich halo-like component, the Splash, formed by heated $\alpha$-rich thick disk stars \citep[][]{Splash}. 

On retrograde motions, a few other chemo-dynamical substructures have been found: 
Sequoia \citep[][]{Myeong2019},  
Thamnos \citep[][]{Koppelman}, and 
Dynamically Tagged Groups \citep[][]{Yuan2020}.

Besides the Helmi stream \citep[HS; ][]{Helmi1999}, and Sagittarius \citep[e.g.][]{Ibata1994}, 
a few prograde streams were discovered at high $z$ from the Galactic plane: Aleph and Wukong \citep[][]{Naidu2020}. 

Much more challenging is the detection of merger debris in the Galactic disk even when chemical and dynamical informations are provided \citep[][]{Helmi2006,Ruchti,Kordopatis2020}. 

Several studies identified stars with kinematics typical of the disks and [Fe/H]$< -1$ \citep[]{Norris, Chiba, Sestito2019, Sestito2020, Dovgal}, even down to 
[Fe/H]$\simeq-6$ according to \citet[][]{DiMatteo}.
However, the origin of these very metal-poor stars ([Fe/H] $<-2.0$) is still a matter of debate. 

This intermediate population, 
with disk kinematics and halo metallicity, 
has been usually considered the tail of the thick disk and dubbed the {\it metal-weak thick disk} \citep[MWTD;][]{Morrison, Kordopatis2013, Beers+2014, Gonzalez}. Furthermore, \citet[][]{Feltzing} associate these objects to a primordial Galactic disk formed {\it in situ} from a chemically unevolved medium before the formation of the canonical thick disk.  

Recently, \citet[][]{Fernandez-Alvar} and \citet[]{Nepal} claimed the existence of an ancient population formed by metal-poor stars and thin disk kinematics. 
The latter authors argued that this population represents the remnant of the primordial thin disk that would have been formed in an ``inside out" manner less than 1 Gyr from Big Bang.

According to other studies, the metal-poor stars with disk orbits may be the tail of a prograde component of the halo, possibly originated by bar resonances or accreted building blocks \citep[][]{BelokurovKravtsov, Dillamore, AnkeAA, Zhang}.

The merging scenario is supported by \citet[][]{ReFiorentin2021} that reported the discovery of five substructures with [Fe/H]$< -1$ and disk kinematics. 
These groups include the Icarus stream, 
whose dynamical properties are consistent with the debris of a dwarf galaxy with a stellar mass of $10^9\, M_{\sun}$ on an initial prograde low-inclination orbit, $\sim10^\circ$.

Also \citet[][]{Mardini} claimed an extragalactic origin of the MWTD component and proposed to refer to as Atari disk. 
This accretion scenario is also validated by high resolution cosmological simulations of MW-like galaxies \citep[][]{Santistevan, Sestito2021, Carollo}. 

Finally, \citet[][]{Malhan2024} reported the existence of two substructures, dubbed ``Shakti" and ``Shiva", with $-2.5 <$ [M/H] $<-1.0$ stars on prograde orbits. However, their origin (in situ field stars trapped by resonances of the rotating bar or remnants of accreted building blocks) is still uncertain. 

In order to discriminate between the scenarios, 
large unbiased (non-kinematically selected) samples of stars with accurate 6D phase-space information and 
chemical properties for classification and 
characterization can be obtained from high-precision data already (or soon to be) available. 
The {\it Gaia} third Data Release \citep[DR3;][]{GaiaCollab.2023:Vallenari} contributes unprecedented accurate measurements of parallax and proper motion for more than $1.4~\rm{billion}$ stars across the whole sky; 
the Apache Point Observatory Galactic Evolution Experiment \citep[APOGEE~DR17;][]{Majewski2017, APOGEEdr17} 
and the Galactic Archaeology with HERMES \citep[GALAH~DR3;][]{DeSilva2015,Buder+2021} have provided high-resolution ($R \sim 22\,500$ in the near-infrared for 733\,901 stars and $R \sim 28\,000$ in the optical for 588\,571, respectively) spectra yielding precise line-of-sight velocities, stellar parameters and abundances for more than $20$ chemical elements. 

Here, we exploit the excellent synergy between the {\it Gaia}, APOGEE and GALAH, and take advantage of these high-quality data to study chemo-kinematical signatures in the local volume, with particular attention to finding and characterizing accreted material towards the disk.

The paper is organised as follows: 
In Sect.~\ref{sec:2}, we assemble a sample of metal-poor accreted/unevolved stars up to 3~kpc of the Sun selected from {\it Gaia} complemented with APOGEE and GALAH. In Sect.~\ref{sec:3}, we explore their phase-space distribution and identify Icarus and GSE candidates. In Sect.~\ref{sec:4}, we further examine their chemical properties to validate an accreted origin for such objects. In Sect.~\ref{sec:5} we present the CMD that confirms an old age of the Icarus members. In Sect.~\ref{sec:6} we compare our results to simulations and further examine Icarus debris in the Toomre diagram and in the space of adiabatic invariants. Finally, we summarize and discuss our results with respect to other studies in Sect.~\ref{sec:7}.


\begin{figure*}
	\centering
	\gridline{\fig{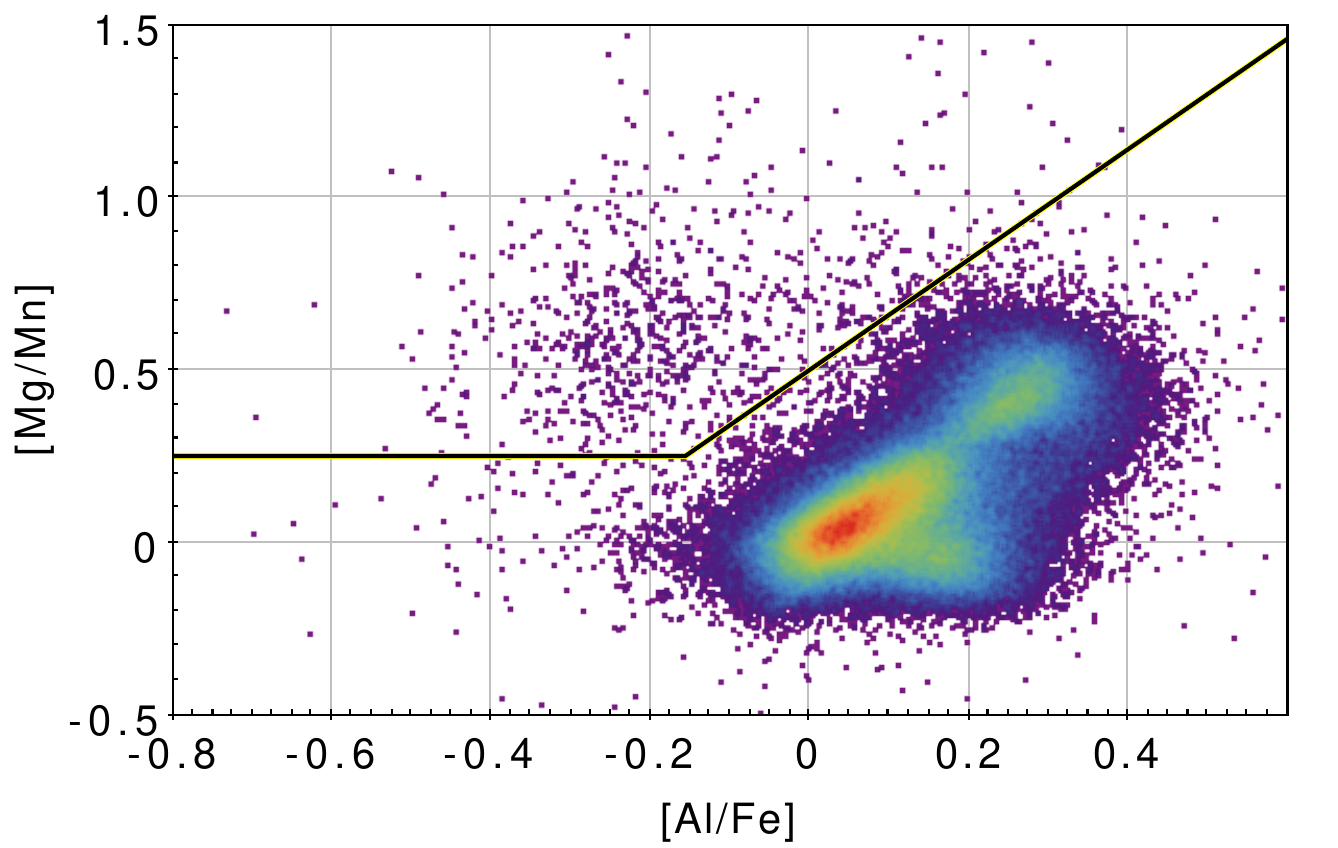}{0.5\textwidth}{}
              \fig{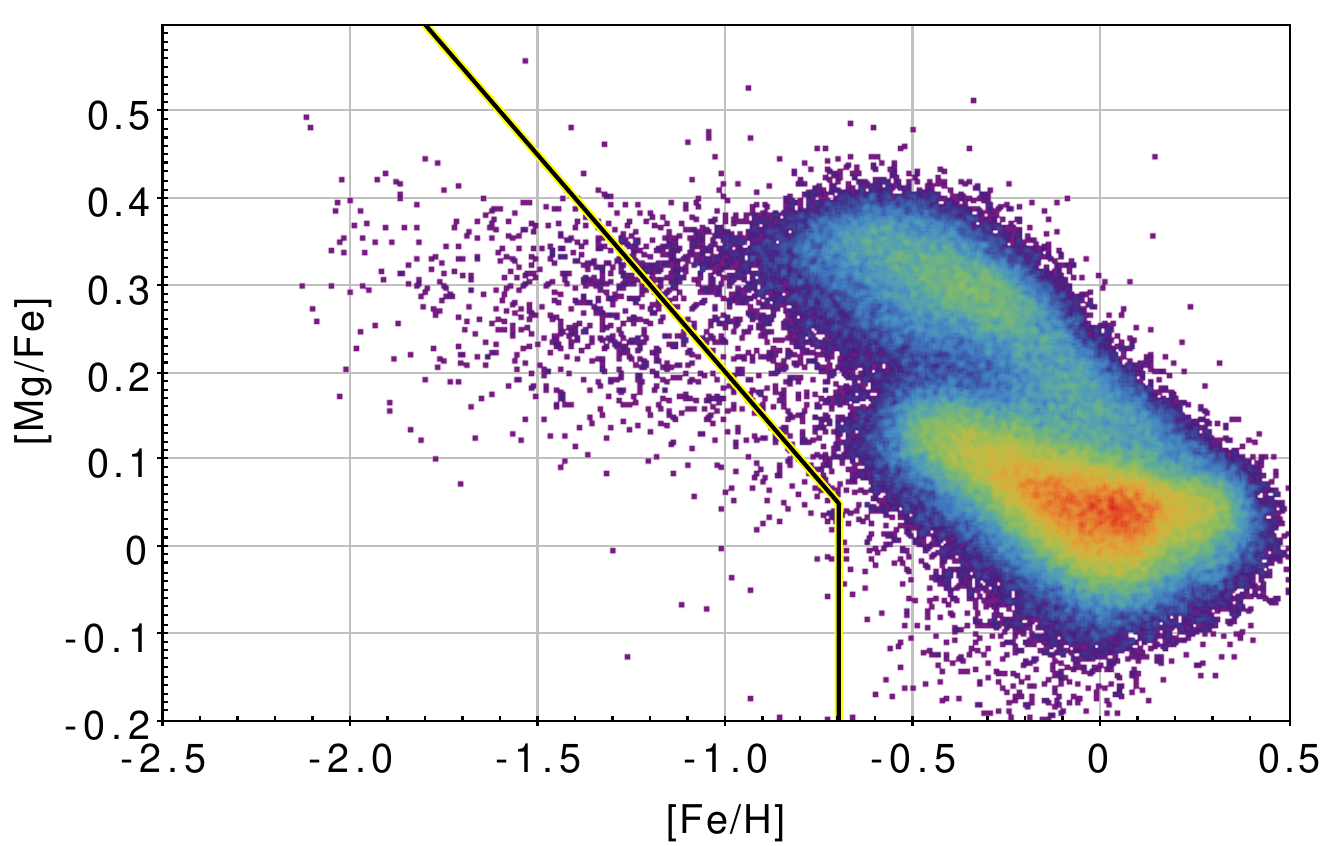}{0.5\textwidth}{}
              }
    \gridline{\fig{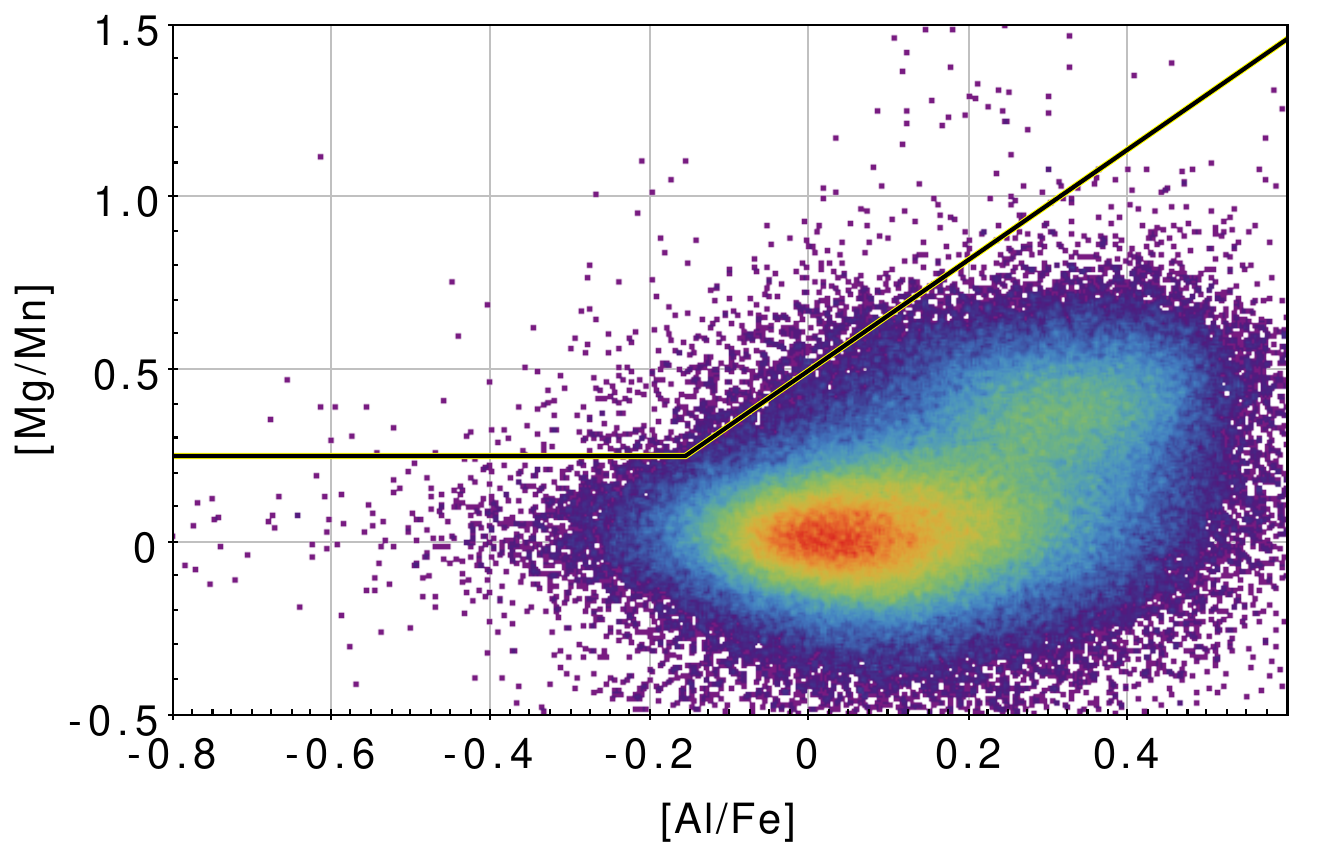}{0.5\textwidth}{}
              \fig{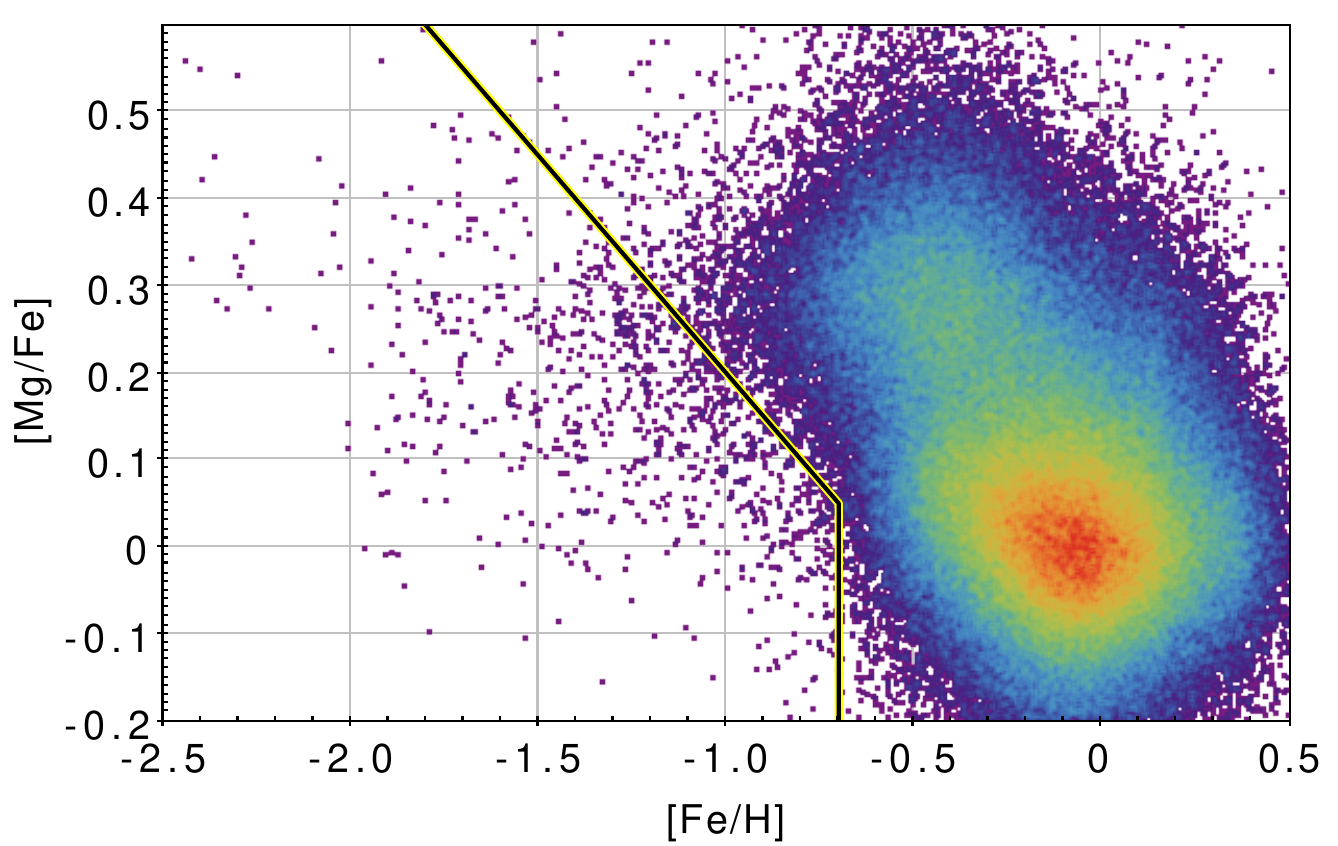}{0.5\textwidth}{}
              }          
	\caption{
	Top panels: chemical distribution, $\rm{[Mg/Mn]}$-$\rm{[Al/Fe]}$ (left) and $\rm{[Mg/Fe]}$-$\rm{[Fe/H]}$ (right), 
	for the $149\,826$ {\it Gaia}-APOGEE stars within $3~\rm{kpc}$ of the Sun. 
	The solid lines represent the locus of accreted/unevolved stars (above/below the lines in the left/right panel, respectively), separated from {\it in situ} populations of thick/thin disk stars. 
	Bottom panels: same distribution as top panels for the $237\,979$ {\it Gaia}-GALAH stars.
          }
	\label{fig:fig1}
\end{figure*}

\begin{figure*}
	\centering
	\gridline{\fig{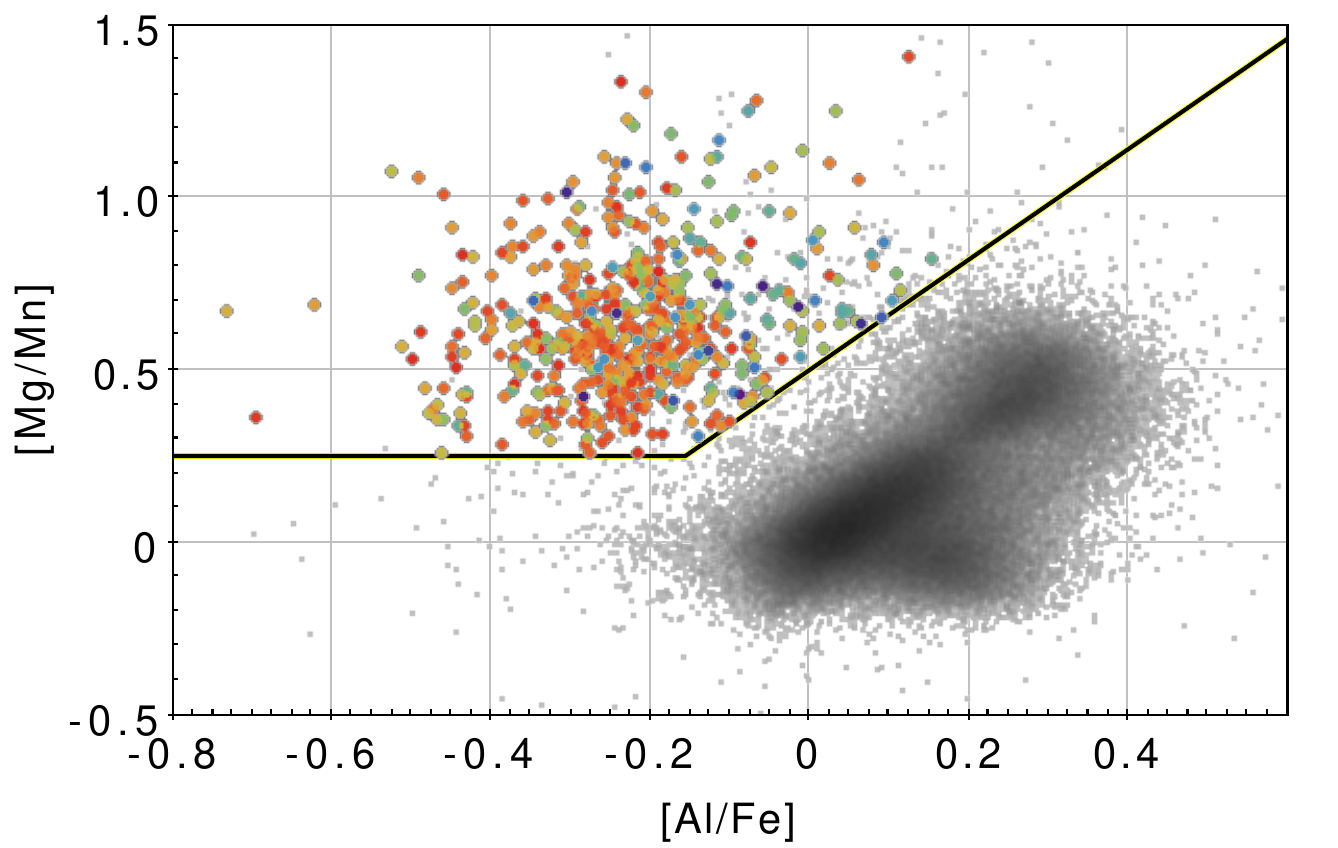}{0.5\textwidth}{}
              \fig{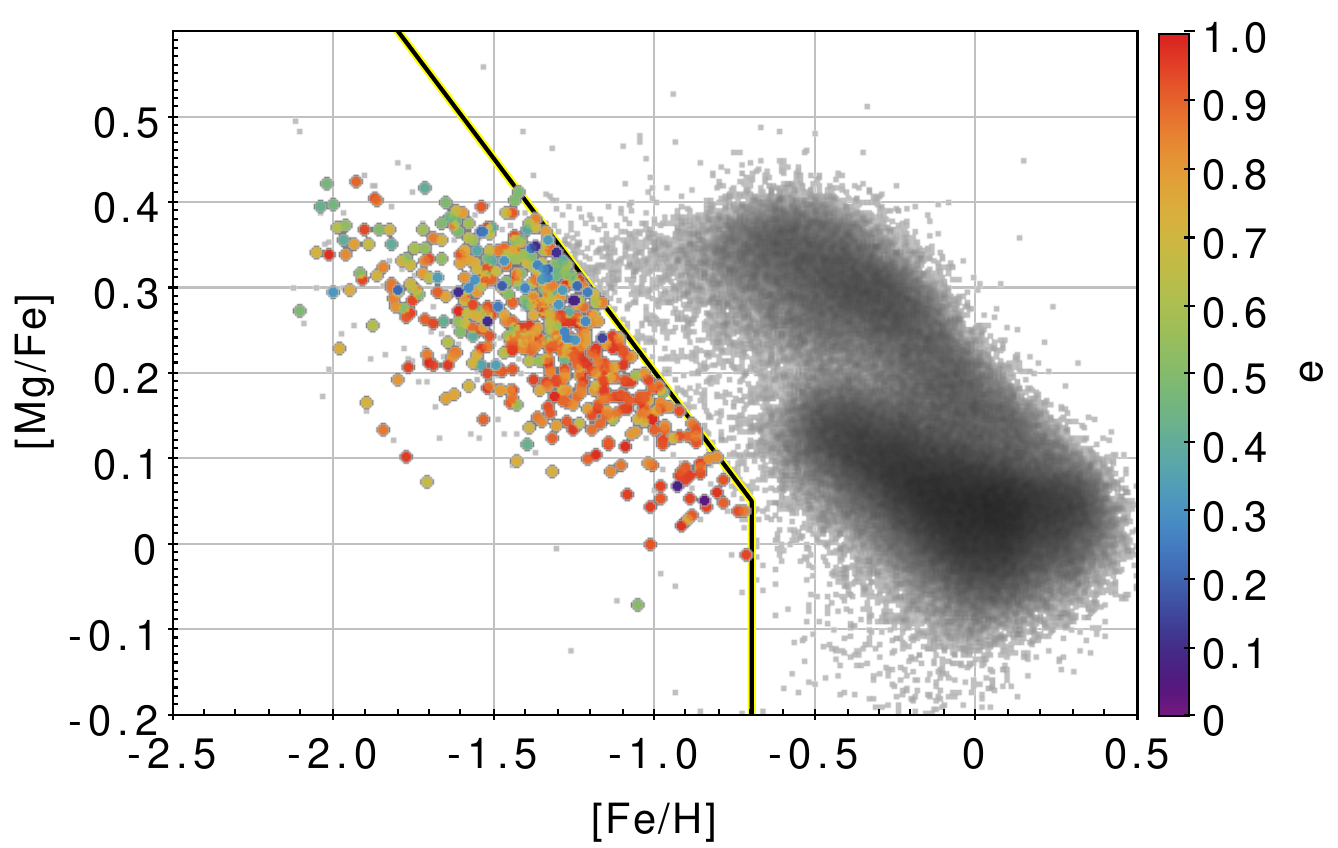}{0.5\textwidth}{}
             }
    \gridline{\fig{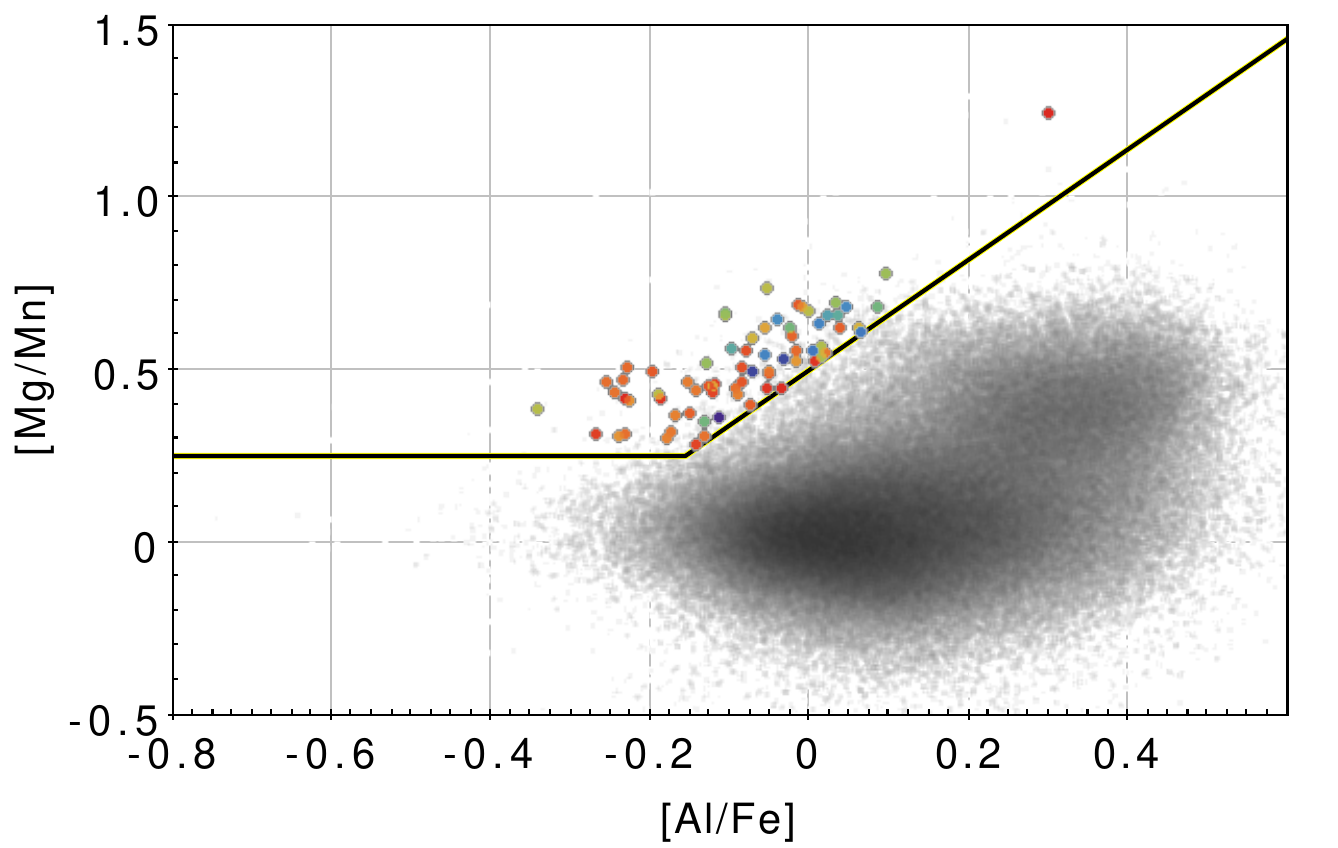}{0.5\textwidth}{}
              \fig{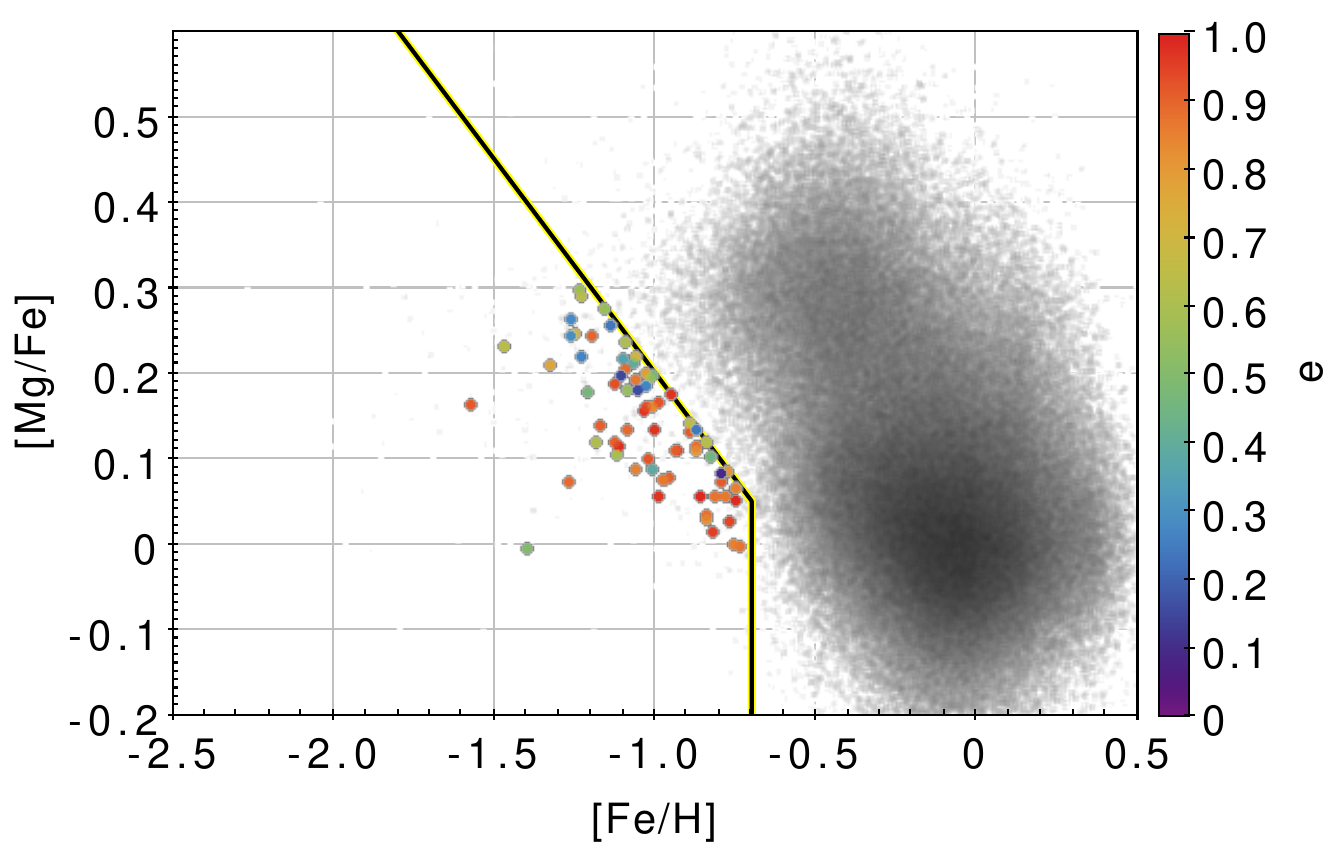}{0.5\textwidth}{}
             }          
	\caption{Top panels:  
	chemical distribution, $\rm{[Mg/Mn]}$-$\rm{[Al/Fe]}$ (left) and $\rm{[Mg/Fe]}$-$\rm{[Fe/H]}$ (right), 
	for the $149\,826$ {\it Gaia}-APOGEE stars. 
	The solid lines represent the adopted selection of $551$ accreted/unevolved stars (color coded according to eccentricity), separated from {\it in situ} thick/thin disk stars (gray). 
	Bottom panels: same distribution as top panels for the $71$ selected stars among the $237\,979$ {\it Gaia}-GALAH stars.
          }
	\label{fig:fig2}
\end{figure*}

\section{Data}\label{sec:2}

For this study, we analyze a chemo-dynamical catalog after cross-matching {\it Gaia} DR3, APOGEE~DR17, and GALAH~DR3. Our catalog includes {\it Gaia} astrometry 
\citep[i.e., sky positions, parallaxes and proper motions; see][]{Lindegren2018a},  
plus line-of-sight velocities and chemical abundances derived with the APOGEE and GALAH stellar spectra parameters pipelines \citep[e. g.,][]{Holtzman2015,Garcia2016,Kos2017,Buder+2021}. 
 
As for the {\it Gaia} data, 
we select objects having Renormalised Unit Weight Error, RUWE, less than $1.4$ in order to discard sources with problematic astrometric solutions, 
astrometric binaries and other anomalous cases \citep[][]{Lindegren2018b}. 
We retain only stars with three {\it Gaia} photometric bands and 
relative parallax error $\varpi/\sigma_\varpi >5$  
to derive trigonometric distances better than $20\%$ as 
\begin{equation}
d=\frac{1}{\varpi - ZP},
\label{eq:1}
\end{equation} 
where $ZP$ 
is the individual parallax zero-point calculated according to \citet[][]{Lindegren2021}.
In addition, {\it Gaia} duplicated sources are excluded from the sample.

For the selection of APOGEE data, 
we remove stars with noisy spectra (\verb+SNR+ $<50$), members of star clusters and dwarf galaxies (\verb+MEMBERFLAG+ $>0$) as well as duplicate observations of a single target 
(\verb+EXTRATARG+ $=16$).  
We also consider only stars
with \verb+STARFLAG+ = 0, indicating no problems in the data reduction. Moreover, we retain stars with no problems in the iron abundance determination (\verb+FE_H_FLAG+ = 0), and in the individual abundances (\verb+X_FE_FLAG+ = 0) for magnesium (Mg), manganese (Mn), and alluminium (Al). 
In case of [Mn/Fe] and [Al/Fe] we also include a few stars with a warning flag on the chemical abundance within 1 grid spacing of grid edge. This sample is further restricted to stars with 
$4000~\rm{K}<T_{\rm{eff}}<6000~\rm{K}$ and $0< \log g < 5$. 

Similarly for the GALAH data,  
we select stars with high signal-to-noise ratio (\verb+snr_c3_iraf+ $> 30$) and with no warning flags on 
the estimated stellar parameters (\verb+flag_sp+ $= 0$), 
iron abundance determination (\verb+flag_fe_h+ $= 0$), 
and individual elemental abundances (\verb+flag_Mg_fe+ $= $ \verb+flag_Mn_fe+ $= $ \verb+flag_Al_fe+ $= 0$), as per \citet[][]{Buder+2021,Buder+2022}. 

In case of objects present in both APOGEE and GALAH, chemical abundances are adopted from APOGEE. 
Therefore, within 3~kpc of the Sun, we have a sample of $387\,805$ stars down to $G\simeq 17~\rm{mag}$.  

Galactic coordinates\footnote{We employ right-handed frames of reference with the axes positive toward the Galactic center, in the direction of Galactic rotation, and toward the North Galactic Pole, respectively.}  
and velocity components are derived by assuming that the Sun is at $z_\sun=0.0208~\rm{kpc}$ and at $8.249~\rm{kpc}$ away from the MW centre,  
and the peculiar velocity of the Sun is $(U,V+V_{\rm LSR},W)_{\sun}=(9.5,250.7,8.56)~\rm{km~s}^{-1}$ 
\citep[][]{Bennett, GravityCollab, Reid}.
Median uncertainties of the resulting Galactocentric velocities are below $0.5~\rm{km~s}^{-1}$ for each component.

We also compute the orbital parameters of each entry (e.g., eccentricity $e$ and the maximum height from the Galactic plane, $z_{\rm max}$) 
by adopting the Galactic potential model \verb+MWPotential2014+ 
from \citet[][]{Bovy2015}
\footnote{We remark that this model assumes a dark matter halo of $0.8\cdot 10^{12}$~M$_{\sun}$. 
Since the estimate of the Galactic mass is still controversial \citep[e.g.,][]{McMillan, Eilers, Jiao, Beordo}, we also validate our findings with a dark matter halo of $1.2\cdot 10^{12}$~M$_{\sun}$ \citep[][]{Bland-Hawthorn} that yields consistent outcomes.}.

Figure~\ref{fig:fig1} shows the chemical planes, $\rm{[Mg/Mn]}$-$\rm{[Al/Fe]}$ and 
$\rm{[Mg/Fe]}$-$\rm{[Fe/H]}$, 
for the full chemo-kinematical catalog. These are useful diagnostic spaces to identify accreted stars in the MW \citep[e.g.,][]{Nissen,Hawkins2015,Das2020,Buder+2024}.

Clearly, these distributions are dominated by thin and thick disk stars. We chemically identify accreted/unevolved stars 
by taking objects with 
\begin{eqnarray}
&&\rm{[Mg/Mn]} >  0.5 + 1.6 \cdot \rm{[Al/Fe]} \label{eq:2}\\
&&\rm{[Mg/Mn]} >  0.25  \label{eq:3}\\
&&\rm{[Mg/Fe]} <  -0.2 - 0.5 \cdot (\rm{[Fe/H]}+0.2) \label{eq:4}\\
&&\rm{[Fe/H]}  <  -0.7  \label{eq:5}
\end{eqnarray}

as for \citet[][]{Horta}, \citet[][]{Feuillet}, and \citet[][]{Mackereth2019}.

In addition, we further remove possible members of nearby globular clusters listed by \citet[][]{GaiaCollab.2018:Helmi}. 
This last selection leaves $622$ accreted/unevolved stars within $3~\rm{kpc}$ of the Sun, that are shown in Figure~\ref{fig:fig2} color coded according to their eccentricity.

\begin{deluxetable}{lrrr}
\tablenum{1}
\tablecaption{\label{table:1} 
Dynamical selection of Icarus and GSE.}
\tablewidth{0pt}
\tablehead{
\colhead{}  & & \colhead{Icarus} &  \colhead{GSE}
}
\startdata
N & & 81& 376\\
$U_0$ &$\rm{(km~s^{-1})}$ & 0 & 0\\
$s_U$ &$\rm{(km~s^{-1})}$ & 150 & 400\\
$V'_0$& $\rm{(km~s^{-1})}$ & 200 & 0\\
$s_V$ &$\rm{(km~s^{-1})}$ & 100 & 150\\
$W_0$ &$\rm{(km~s^{-1})}$ & 0 & 0\\
$s_W$ &$\rm{(km~s^{-1})}$ & 150 & 200\\
$e$ & &$-$&  $>0.7$
\enddata
\tablecomments{where $V'_0 = V_0+V_{\rm{LSR}}$
}
\end{deluxetable}


\section{Icarus and GSE Membership}\label{sec:3}

\begin{figure*}
	\centering
	\gridline{\fig{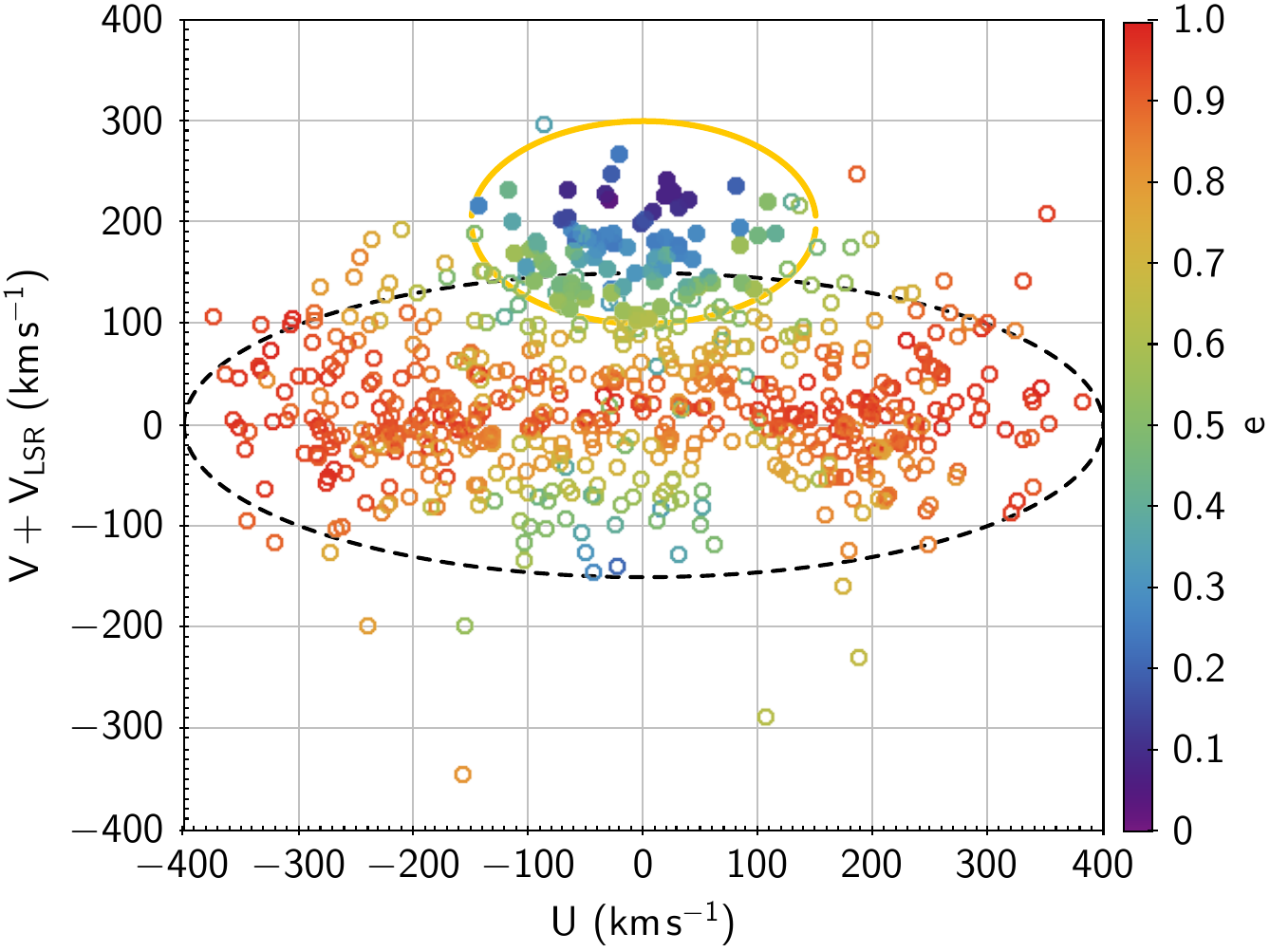}{0.5\textwidth}{}
              \fig{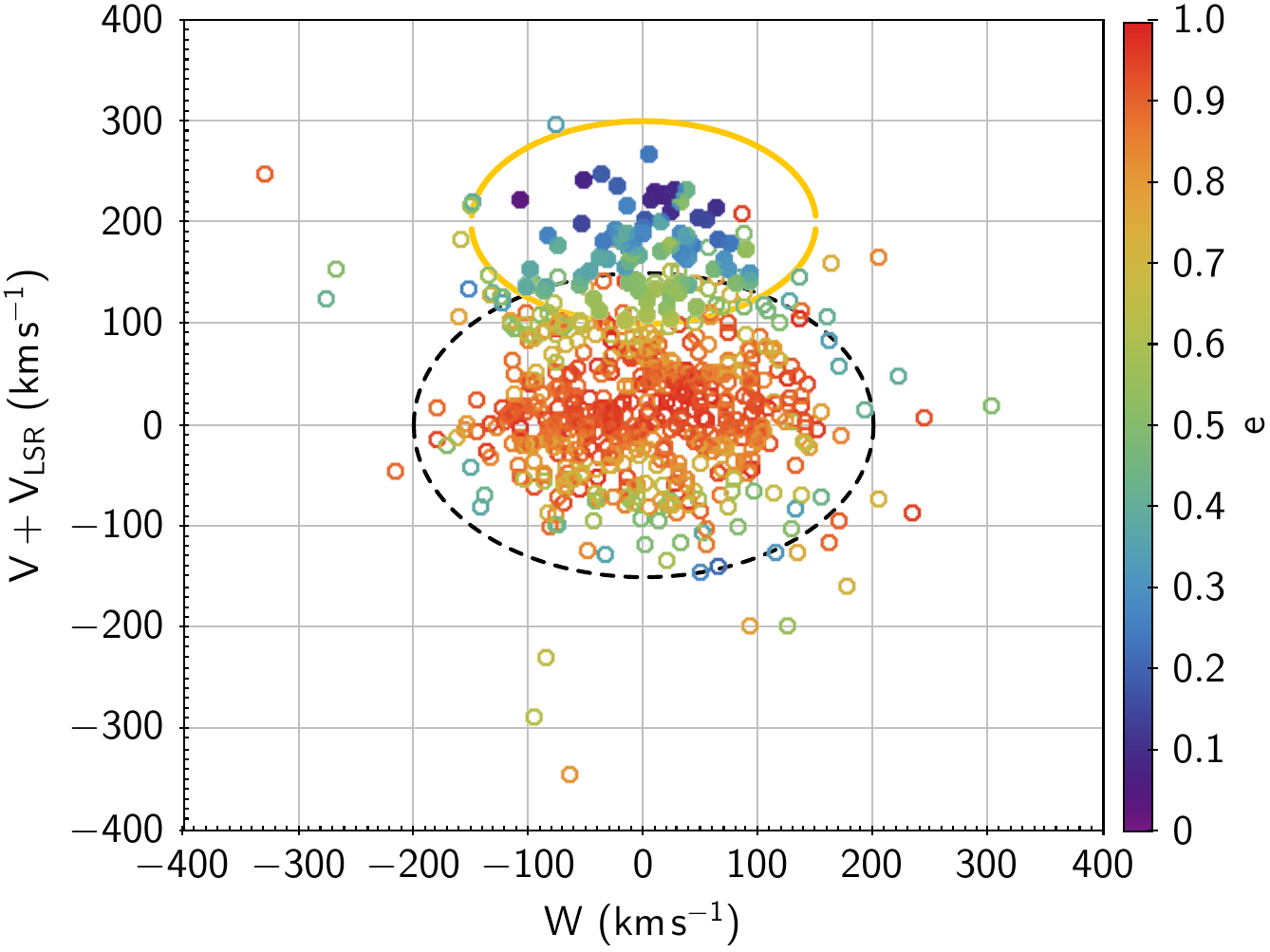}{0.5\textwidth}{}
             }
    \gridline{\fig{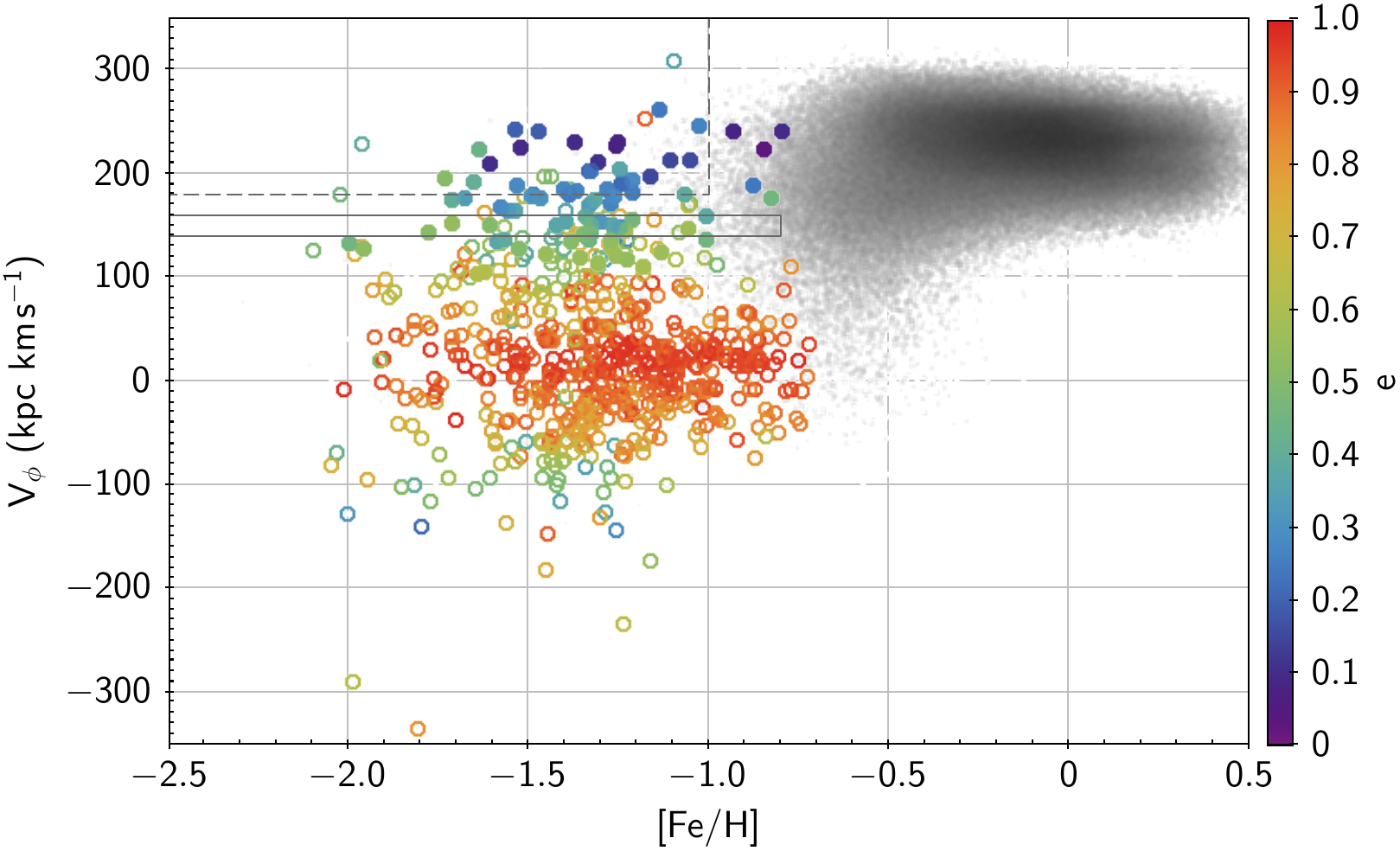}{0.6\textwidth}{}
             }          
	\caption{
	Top panels: 3D velocity space for the 
	$622$ chemically selected nearby stars color coded according to eccentricity. 
	Icarus stars are highlighted as filled dots within the Icarus locus (solid line). 
	The dashed line encircles GSE \citep[][]{Helmi2018,Belokurov2018}. Here we identify 376 objects with $e>0.7$. 
	A few stars from Thamnos  \citep[][]{Koppelman}, Sequoia  \citep[][]{Myeong2019}, and HS  \citep[][]{Helmi1999} are present.
        Bottom panel: the $V_{\phi}$--[Fe/H] distribution of the chemically selected {\it ex situ} sample, as above. 
	In the background the full sample within 3 kpc of the Sun is shown.
  	The chemo-kinematical selection thresholds,  
   $140< V_{\phi}<160~\rm{km~s}^{-1}$ and [Fe/H]$< -1$ 
   for the Atari disk / MWTD by \citet[][]{Mardini} 
   and $V_{\phi}>180~\rm{km~s}^{-1}$ and [Fe/H]$<-0.8$ 
   for the metal-poor thin disk by \citet[][]{Nepal} 
   are represented by the dashed/solid lines, respectively.   
          }
	\label{fig:fig3}
\end{figure*}

Figure~\ref{fig:fig3} 
shows the full accreted/unevolved sample in 3D velocity space (top panel) color coded by eccentricity. 
The sample is dominated by two main structures with distinct dynamical properties: 
halo stars on high eccentricity orbits and mean null rotation, 
$V+V_{\rm{LSR}}\sim~0$, 
and low-eccentricity stars with disk-like kinematics, $(U,V+V_{\rm{LSR}},W)\simeq (0,200,0)$. 
The former corresponds to GSE, the latter includes Icarus and the associated prograde 
Groups~3, 5, 6, 7 
found by \citet[][]{ReFiorentin2021}. 

A few counter-rotating stars from Thamnos \citep[][]{Koppelman}, and Sequoia \citep[][]{Myeong2019} are also visible. 
We select members of Icarus and GSE, by means of the velocity threshold:

\begin{equation}
\left(\frac{U-U_0}{s_U}\right)^2 + \left(\frac{V-V_0}{s_V}\right)^2 + \left(\frac{W-W_0}{s_W}\right)^2 < 1
\label{eq:6}
\end{equation}
where $(U_0,V_0,W_0)$ and $(s_U,s_V,s_W)$ are respectively the center and the principal axes of the triaxial velocity ellipsoid. 
Here, we assume the 
parameters listed in Table~\ref{table:1} that   
define the locus of Icarus and GSE shown in Figure~\ref{fig:fig3} 
as the solid and dashed lines, respectively.  
Thus, we identify 81 members of Icarus (filled circles in Figure~\ref{fig:fig3}) and 376 GSE members with the additional constrain $e> 0.7$. The mean dynamical and chemical properties of Icarus and GSE are reported in Table~\ref{table:2} and Table~\ref{table:3}.

The bottom panel of Figure~\ref{fig:fig3} shows the Galactic rotational velocity $V_\phi$ vs.\ [Fe/H] diagram for all the 622 accreted/unevolved chemically selected objects; 
the full sample of 414\,098 high quality astrometric and spectroscopic stars within 3~kpc of the Sun is overplotted (gray dots). 
The solid lines include the region with 
[Fe/H]$< -0.8$ and $140< V_{\phi}<160~\rm{km~s}^{-1}$ 
that defines the Atari disk / MWTD \citep[][]{Mardini}. 
The dashed lines show the region with 
[Fe/H]$< -1$ and $V_{\phi}>180~\rm{km~s}^{-1}$ that, 
combined with $z_{\rm max}<1$~kpc, defines 
the metal-poor thin disk according to \citet[][]{Nepal}.

We point out that the kinematical distribution of the Atari disk and the metal-poor thin disk overlap with our Icarus group (see discussion in Sect.~\ref{sec:7}).

\begin{deluxetable*}{lrrrrrrrrrrcccc}
\tablenum{2}
\label{table:2}
\tablecaption{Dynamical Properties (mean and dispersion) of Icarus and GSE.}
\tablewidth{0pt}
\tablehead{
\colhead{Group}  & 
\multicolumn{2}{c}{$U$} & 
\multicolumn{2}{c}{$V+V_{\rm{LSR}}$} & 
\multicolumn{2}{c}{$W$} & 
\multicolumn{2}{c}{$L_{z}$} & 
\multicolumn{2}{c}{$L_{xy}$} & 
\multicolumn{2}{c}{$z_{\rm max}$}  &
\multicolumn{2}{c}{$e$}\\
& 
\multicolumn{2}{r}{$\rm{(km~s^{-1})}$} & 
\multicolumn{2}{r}{$\rm{(km~s^{-1})}$} & 
\multicolumn{2}{r}{$\rm{(km~s^{-1})}$} & 
\multicolumn{2}{c}{$\rm{(kpc~km~s^{-1})}$} & 
\multicolumn{2}{c}{$\rm{(kpc~km~s^{-1})}$} & 
\multicolumn{2}{c}{$\rm{(kpc)}$} &  & \\
& 
\multicolumn{1}{c}{$\mu$} & \multicolumn{1}{c}{$\sigma$}&  
\multicolumn{1}{c}{$\mu$} & \multicolumn{1}{c}{$\sigma$}& 
\multicolumn{1}{c}{$\mu$} & \multicolumn{1}{c}{$\sigma$}& 
\multicolumn{1}{c}{$\mu$} & \multicolumn{1}{c}{$\sigma$}&  
\multicolumn{1}{c}{$\mu$} & \multicolumn{1}{c}{$\sigma$}& 
\multicolumn{1}{c}{$\mu$} & \multicolumn{1}{c}{$\sigma$}&  
\multicolumn{1}{c}{$\mu$}& \multicolumn{1}{c}{$\sigma$}
}
\startdata
Icarus & $-13$&$59$&$171$&$37$&$3$& $46$& $1340$ & $307$ & $366$ & $212$ & $1.74$ & $1.06$ &  $0.36$ & $0.16$ \\
GSE & $-9$ & $187$ & $10$ & $43$ &$2$ & $70$ & $97$ & $332$ &$455$&$292$ & $4.95$ & $3.97$ &  $0.88$ & $0.07$ \\
\enddata
\end{deluxetable*}

\begin{deluxetable*}{lrrrrrrrrrrrrrrrrrrrrrrrrrrr}
\tablenum{3}
\label{table:3}
\tablecaption{Chemical Properties (mean and dispersion) of Icarus and GSE.}
\tablewidth{0pt}
\tablehead{
\colhead{Group}  & 
\multicolumn{2}{c}{$\rm{[Fe/H]}$}  & 
\multicolumn{2}{c}{$\rm{[Mg/Fe]}$} & 
\multicolumn{2}{c}{$\rm{[Al/Fe]}$} & 
\multicolumn{2}{c}{$\rm{[Mn/Fe]}$} & 
\multicolumn{2}{c}{$\rm{[(C+N)/O]}$} & 
\multicolumn{2}{c}{$\rm{[Ni/Fe]}$} & 
\multicolumn{2}{c}{$\rm{[Y/Eu]}$} & 
\multicolumn{2}{c}{$\rm{[Eu/Mg]}$} & 
\multicolumn{2}{c}{$\rm{[Ba/Y]}$} &
\multicolumn{2}{c}{$\rm{[Ba/Mg]}$} \\
& 
\multicolumn{1}{c}{$\mu$}& \multicolumn{1}{c}{$\sigma$}& 
\multicolumn{1}{c}{$\mu$} & \multicolumn{1}{c}{$\sigma$}& 
\multicolumn{1}{c}{$\mu$} & \multicolumn{1}{c}{$\sigma$}& 
\multicolumn{1}{c}{$\mu$} & \multicolumn{1}{c}{$\sigma$}& 
\multicolumn{1}{c}{$\mu$} & \multicolumn{1}{c}{$\sigma$}& 
\multicolumn{1}{c}{$\mu$} & \multicolumn{1}{c}{$\sigma$}& 
\multicolumn{1}{c}{$\mu$} & \multicolumn{1}{c}{$\sigma$}& 
\multicolumn{1}{c}{$\mu$} & \multicolumn{1}{c}{$\sigma$}& 
\multicolumn{1}{c}{$\mu$} & \multicolumn{1}{c}{$\sigma$}& 
\multicolumn{1}{c}{$\mu$} & \multicolumn{1}{c}{$\sigma$}
}
\startdata
Icarus & $-1.35$ & $0.24$ & $0.27$ & $0.07$ &$-0.13$ & $0.13$ & $-0.39$ & $0.16$ & $-0.43$ &$0.35$&$-0.004$&$0.034$&$-0.20$&$0.20$&$0.17$&$0.14$&$0.12$&$0.21$&$0.10$&$0.33$\\
GSE    & $-1.27$ & $0.28$ & $0.21$ & $0.09$ &$-0.23$ & $0.11$ & $-0.39$ & $0.18$ & $-0.36$ &$0.33$&$-0.054$&$0.059$&$-0.37$&$0.22$&$0.34$&$0.18$&$0.19$&$0.23$&$0.17$&$0.35$\\
\enddata
\tablecomments{In case of stars present in APOGEE and GALAH, [Al/Fe], [Mg/Fe], and [Mn/Fe] are from APOGEE.\\
For Icarus, mean and dispersion of [Y/Eu] and [Ba/Mg] are derived excluding two chemical outliers with [Y/Eu]$ >+0.2$ and [Ba/Mg]$ >+1$ (see text).}
\end{deluxetable*}

\begin{figure*}
	\centering
	\gridline{\fig{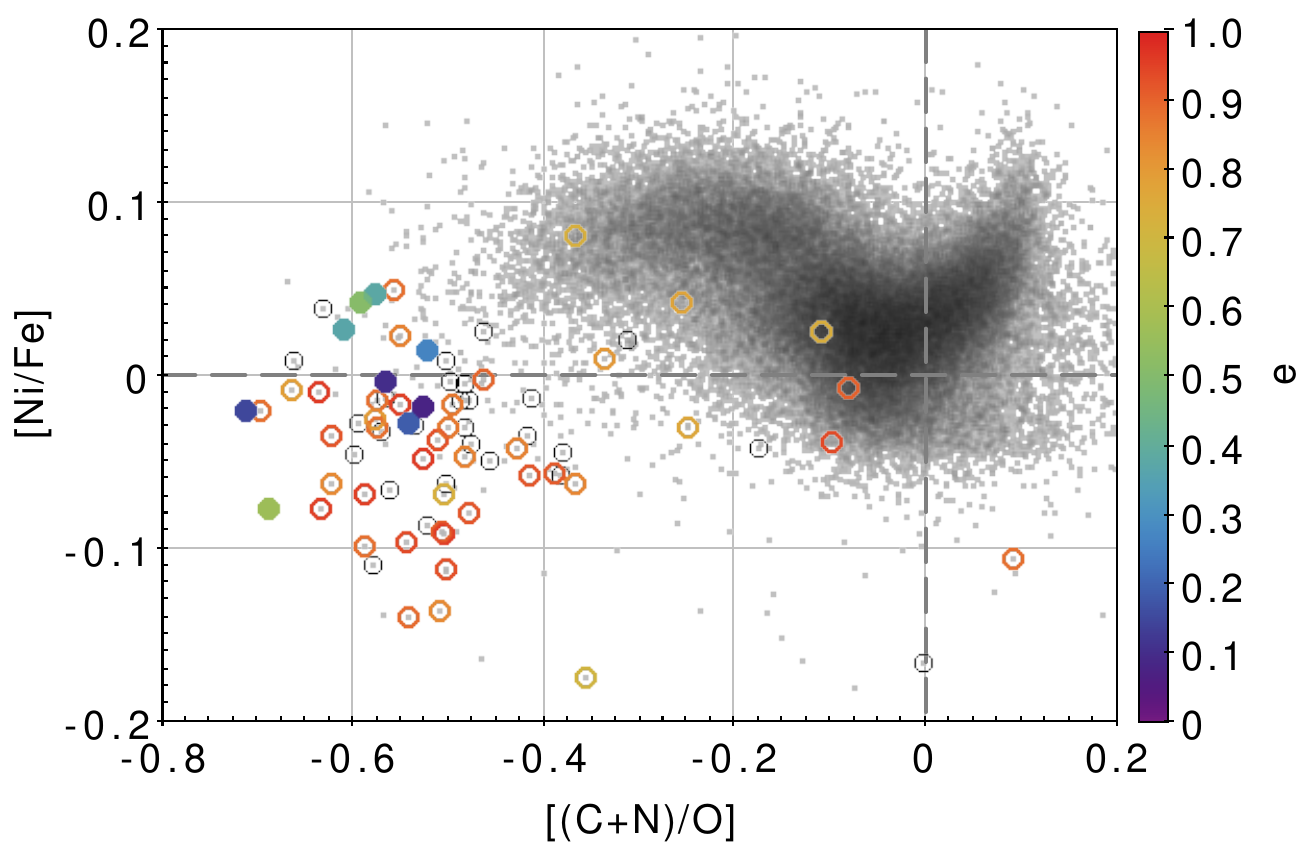}{0.7\textwidth}{}
             }
	\caption{Chemical validation of Icarus. The [(C+N)/O]--[Ni/Fe] plane for the subset of high quality APOGEE giants (gray dots). 
	Icarus and GSE members are shown respectively as filled and open circles, color coded by eccentricity. 
	These are highlighted among the chemically selected accreted/unevolved stars (black open circles).
          }
	\label{fig:fig4}
\end{figure*}

\begin{figure*}
	\centering
	\gridline{\fig{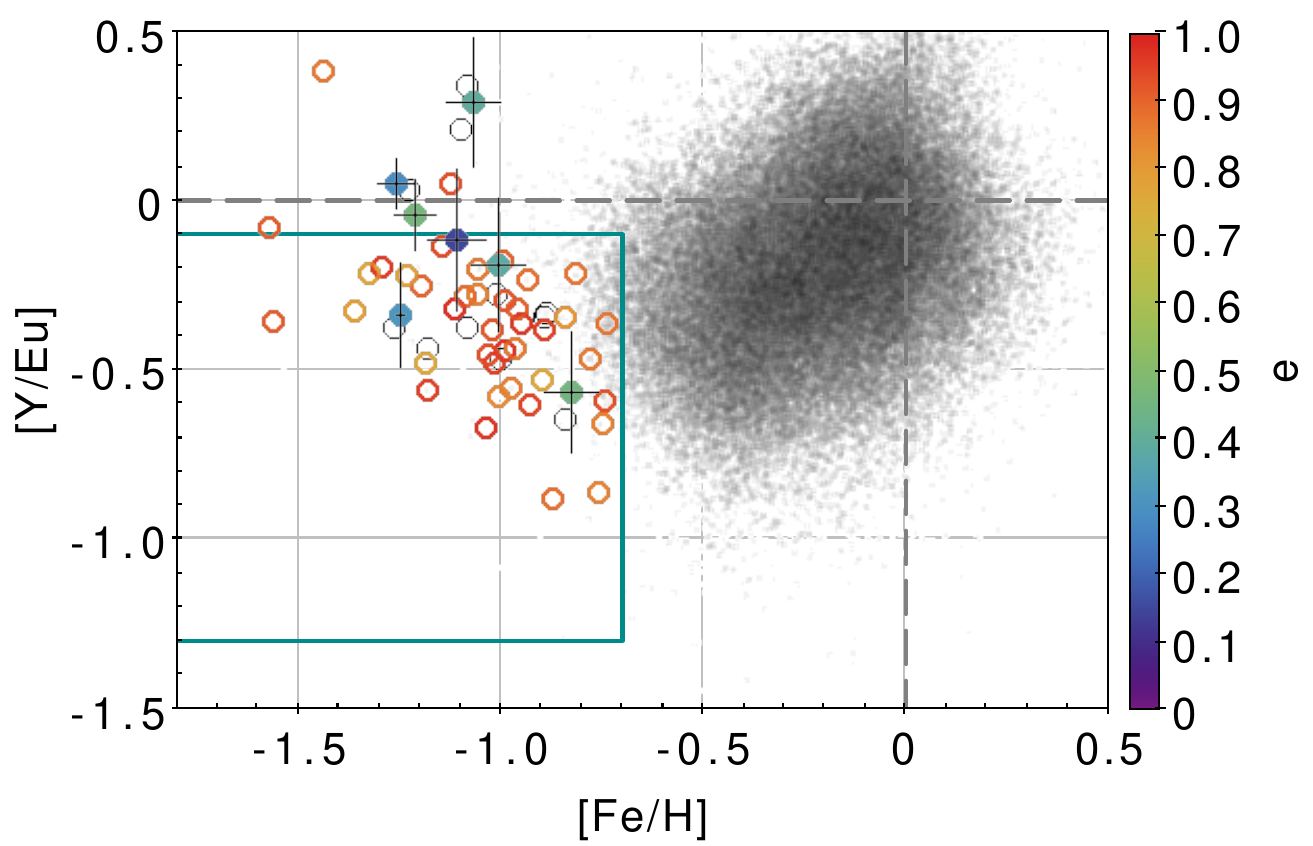}{0.5\textwidth}{}
              \fig{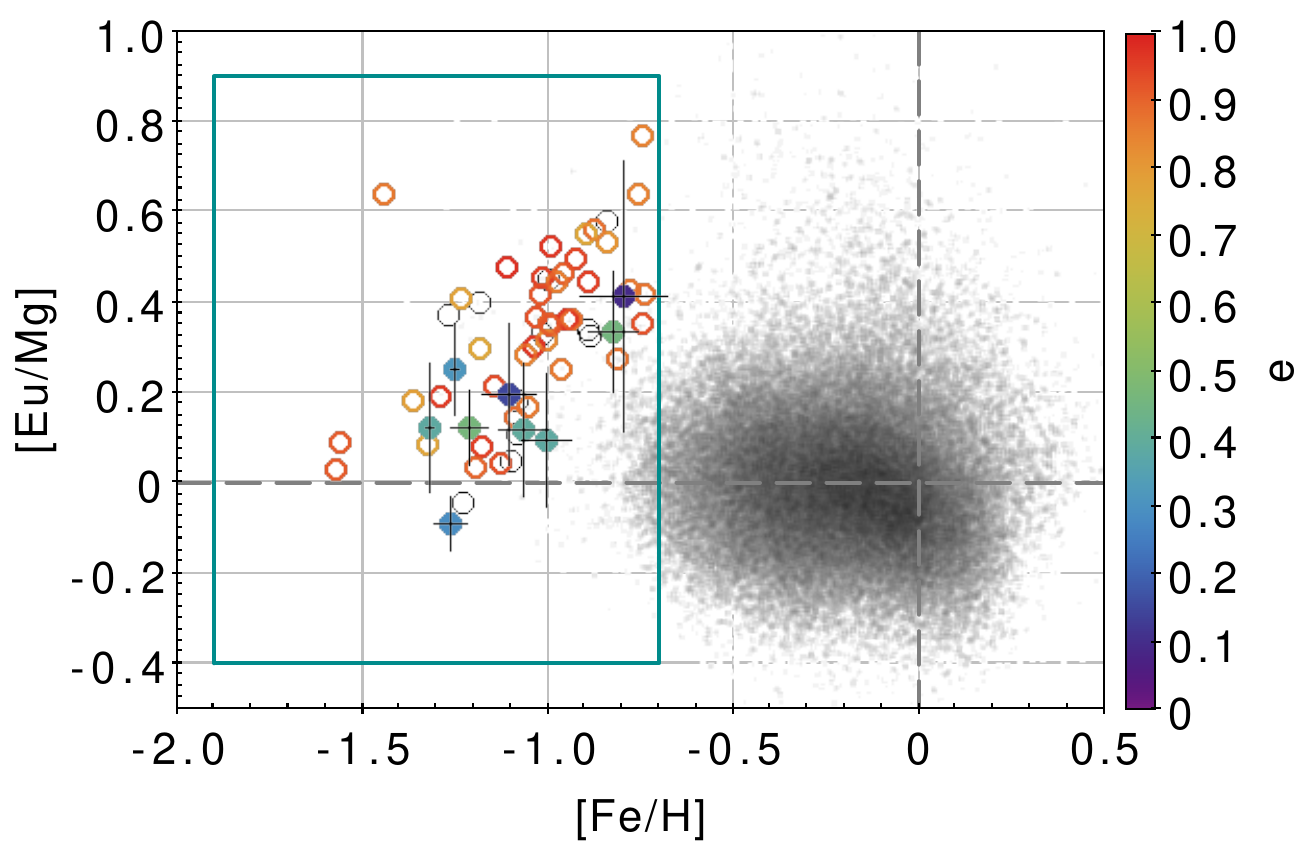}{0.5\textwidth}{}
              }
    \gridline{\fig{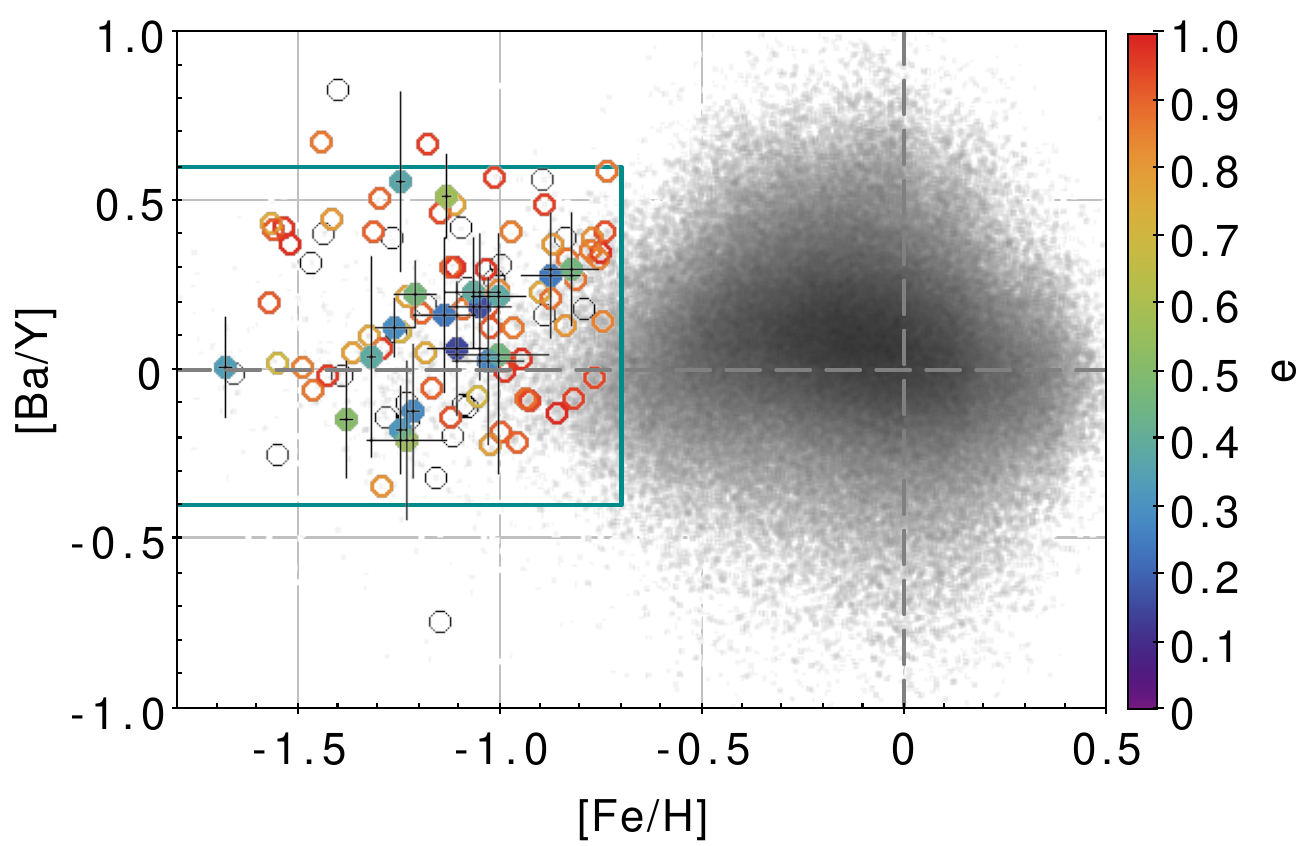}{0.5\textwidth}{}
              \fig{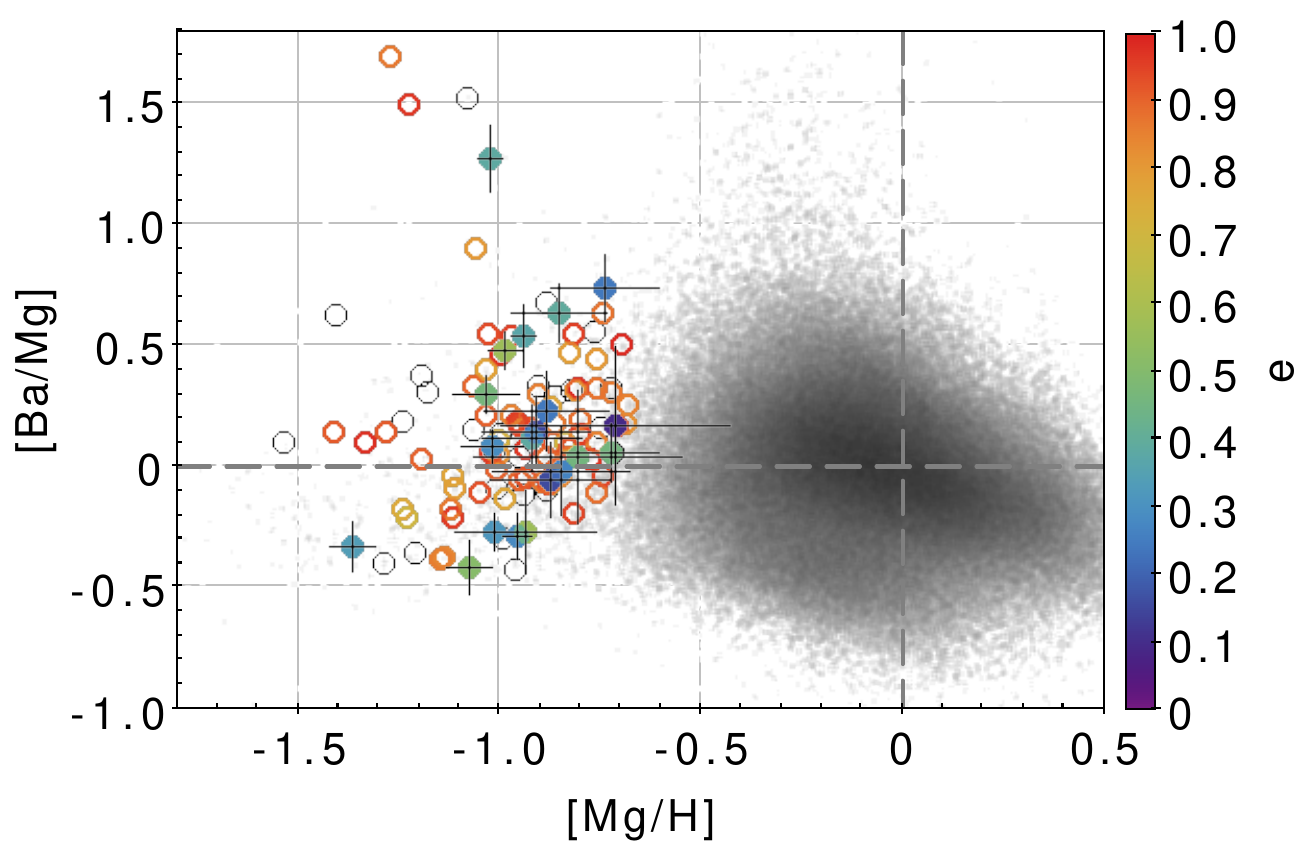}{0.5\textwidth}{}
              }
	\caption{Chemical validation of Icarus. 
    Top panels: 
	the [Y/Eu]--[Fe/H] and [Eu/Mg]--[Fe/H] distributions are shown for the subset of sources that have estimates of 
	europium and yttrium abundance relative to iron without warnings from GALAH (left and right, respectively). 
	Icarus, GSE, and the remaining chemically accreted/unevolved members are represented as in Figure~\ref{fig:fig4}.
	Solid boxes define the ``accreted regions" that enclose the low-$\alpha$ stars from \citet[][]{Nissen} and the ``blob stars" reported by \citet[][]{Carrillo}.
	Bottom-left panel:  as above, but the [Ba/Y]--[Fe/H] plane for all the sources with GALAH estimates of barium and yttrium abundance relative to iron with no warnings. 
	Bottom-right panel: the [Ba/Mg]--[Mg/H] plane that identifies the contributions from AGB stars with respect to SNe II. In particular, for [Ba/Mg] $\ga 0.9$, the AGB contribution is dominant.
          }
	\label{fig:fig5}
\end{figure*}

\section{Chemical validation}\label{sec:4}
In order to better understand the origin of the old and metal-poor stars with disk kinematics belonging to Icarus, we explore other chemical spaces where stars from accreted dwarf galaxies can be distinguished from the Galactic {\it in situ} populations \citep[e.g.,][]{Montalban, Carrillo}.

Figure~\ref{fig:fig4} shows the nickel abundance relative to iron [Ni/Fe] and the sum of carbon and nitrogen abundance relative to oxygen [(C+N)/O] for the giants ($\log g < 2.8$) with no warnings on nickel, carbon, nitrogen, and oxigen from the APOGEE survey\footnote{\tt NI\_FE\_FLAG = C\_FE\_FLAG = N\_FE\_FLAG = O\_FE\_FLAG = 0}. 
Members of Icarus (filled circles) and GSE (open circles) are color coded according to eccentricity. Other stars from  our chemically accreted/unevolved sample are shown as black open circles. From this figure it is clear that Icarus and GSE samples lie at the lowest values of [Ni/Fe], and are clearly separated from the rest of the MW stars (gray dots).

This result is consistent with an {\it ex situ} origin for GSE and the few outliers located within the MW distribution may be due to systematic errors affecting the APOGEE abundances or to contamination by heated disk stars. In fact, 
several studies have linked stars with lower [Ni/Fe] ratios to an accreted origin from dwarf satellite galaxies \citep[e.g.,][]{Nissen, Hawkins2015, Montalban}. This lower [Ni/Fe] ratio can be attributed to the mass-dependence of yields from Type II Supernovae \citep[SNe II;][]{Chieffi, Andrews}. 
Specifically, lower-mass SNe II contribute lower abundances of elements such as Mn, cobalt (Co), and Ni. This deficiency reflects a different star formation history typical of dwarf systems, as it was already observed in satellite galaxies such as Sagittarius \citep[][]{Hasselquist,Sestito2024b}, 
Fornax \citep[][]{Lemasle},
and the LMC \citep[][]{VanderSwaelmen}. 

At the same time a low [(C+N)/O] points towards a top-light Initial Mass Function (IMF) scenario, which is again connected to the mass dependence of Type II SN yields, where C is more mass-dependent than O. This low ratio may indicate a depletion of these light elements in the primordial gas from which the stars originated. The lower levels of carbon in the MW satellites compared to the MW halo have been widely discussed in both observational \citep[][]{Lucchesi, Sestito2024c} and theoretical studies \citep[][]{Kobayashi, Vanni}. These findings have been associated with variations in the capacity of different systems to retain supernova ejecta. Carbon production is dependent on the energy and yield of SNe II, with more energetic supernovae producing lower [C/Fe] ratios compared to less energetic ones. We estimate a mean abundance of $\langle$[C/Fe]$\rangle\simeq 0$ for our subset of Icarus and GSE members with no warning on [C/Fe]. As shown in \citet[][see left panel of Fig.~5]{Sestito2024c}, 
these values correspond to the mean abundance of MW halo stars in the same metallicity range of our sample, [Fe/H] $\simeq -1.4$. This corroborates the hypothesis that the low carbon fraction observed in Icarus aligns with the chemical evolution of a LMC-like dwarf galaxy with a stellar mass of $10^9~M_\sun$, similar to what estimated for GSE.

Recent studies claim that low metallicity stars with disk kinematics represent the primordial disk \citep[][]{Feltzing, Fernandez-Alvar, Nepal}. However, the significant overlap between the Icarus and GSE members in this chemical space seems to support the hypothesis of an {\it ex situ} origin for all these stars, in particular for the sources with [Ni/Fe]$ <0$ \citep[see][]{Montalban, Ortigoza-Urdaneta}. 

To further validate our hypothesis, we look into the neutron (n)-capture abundance ratios in order to understand which processes dominate the chemical evolution of our sample of accreted/unevolved stars. 

In particular, as an indicator of the relative contributions of slow (s)- and rapid (r)-processes in the creation of these elements, 
we explore the distribution of [Y/Eu], [Eu/Mg] and [Ba/Y] as a function of [Fe/H] for stars with GALAH estimates with no warning flags on yttrium, europium and barium (see Figure~\ref{fig:fig5}, top and bottom-left panels). 
The error bars of the Icarus members are derived from uncertainties on the chemical abundances given by the GALAH and APOGEE surveys, ranging from 0.08 to 0.15 for [Mg/Fe] and [Y/Fe], respectively. Solid boxes contour the corresponding regions in Figs. 10-11 of \citet[][]{Carrillo} containing GSE stars selected by these authors and complemented with low-$\alpha$ stars from \citet[][]{Nissen}. 

With respect to the accreted regions, the similarity in chemistry of our Icarus and GSE stars is apparent. Indeed, very few stars fall outside the accreted boxes. Icarus members show a remarkable spread in n-capture elements which can be indicative that both r- and s-process channels (respectively responsible of Eu, and Y, Ba) are present in the chemical enrichment history of the system. Moreover, the spread in heavy-element abundances, particularly at the very metal-poor end of the MW's stellar population, remains an unresolved issue in chemical evolution models. One possibility that could account for this diversity is the varied contribution of accreted satellites to the MW's stellar composition. This idea aligns with the scenario presented in this work.

From the various panels it is interesting to notice that the level of Eu is comparable with the ones presented by GSE stars. Many studies reported an enhancement in r-process elements in GSE and dwarf satellites compared to the MW \citep[e.g.,][]{Aguado, Matsuno, Naidu2022}. This high level is usually connected to neutron star merger, which can act on delayed time scales \citep[][]{Skuladottir2020, Matsuno, Naidu2022}.

The [Ba/Y] ratio aligns with observations from both accreted and {\it in situ} MW stars (see Figure~\ref{fig:fig5}, bottom-left panel). This ratio serves as a diagnostic for the relative contributions of light s-process elements (e.g., Y) and heavy s-process elements (e.g., Ba). These are dubbed first-peak and second-peak n-capture elements, respectively, and are primarily produced by asymptotic giant branch (AGB) stars. However, different masses of AGB stars lead to varying levels of first-peak and second-peak element formation \citep[][]{Travaglio, Kobayashi, Goswami, Pignatari2010}. In our case, it appears that both low-mass and intermediate-mass AGB stars may have contributed to the production of Ba and Y, respectively.

The bottom-right panel of Figure~\ref{fig:fig5} shows the chemical space [Ba/Mg] vs.\ [Mg/H] that is a useful proxy for identifying contributions from AGB stars. Metal-poor AGB stars are expected to increase the [Ba/Mg] ratio through s-process enrichment \citep[e.g.,][]{Pignatari2008, Cescutti, Sestito2024}, leading to an increase in this ratio as [Mg/H] rises. In contrast, if only SNe II contribute, the ratio would remain flat \citep[][]{Cowan}. 

Our Icarus stars seem to show this increasing trend that has been also observed in other dwarf systems, 
such as Fornax \citep[][]{Letarte}, Ursa Minor \citep[][]{Sestito2023}, and Sagittarius \citep[][]{Sestito2024b}.

Finally, we point out the presence of a few chemical outliers in the top-left and bottom-right panels of Figure~\ref{fig:fig5}, such as the two Icarus stars\footnote{{\it Gaia} Source Ids: 4047122020635707520, 2622023937311599104}
with  [Y/Eu] $ >0$ and [Ba/Mg] $ >1$. These stars are also enriched in heavy elements (e.g.\ [Ba/Fe] $>0.8$), which could suggest either binary star systems or different enrichment processes.  The latter hypothesis is favored, since there are no hints of binarity in the main {\it Gaia} catalog and in the non-single stars {\it addendum}, nor in the multi-band photometry from {\it Gaia}, 2MASS \citep[][]{Cutri2003}, and AllWISE \citep[][]{Cutri2014},
that we used to model the spectral energy distributions (SEDs) of these two sources.

In conclusion, the overlap between Icarus and GSE stars supports an accreted origin for both populations. The fact that the chemical distributions are not exactly the same is consistent with an origin from two progenitors with different chemo-dynamical histories.
However, due to the limited statistics of our targets and the constrained chemical space, we cannot entirely rule out the possibility of a primordial disk origin. Higher resolution follow-up observations are needed to further clarify the origin of these stars.


\section{Age estimate}\label{sec:5}

\begin{figure*}
   \centering
    \includegraphics[width=0.7\linewidth]{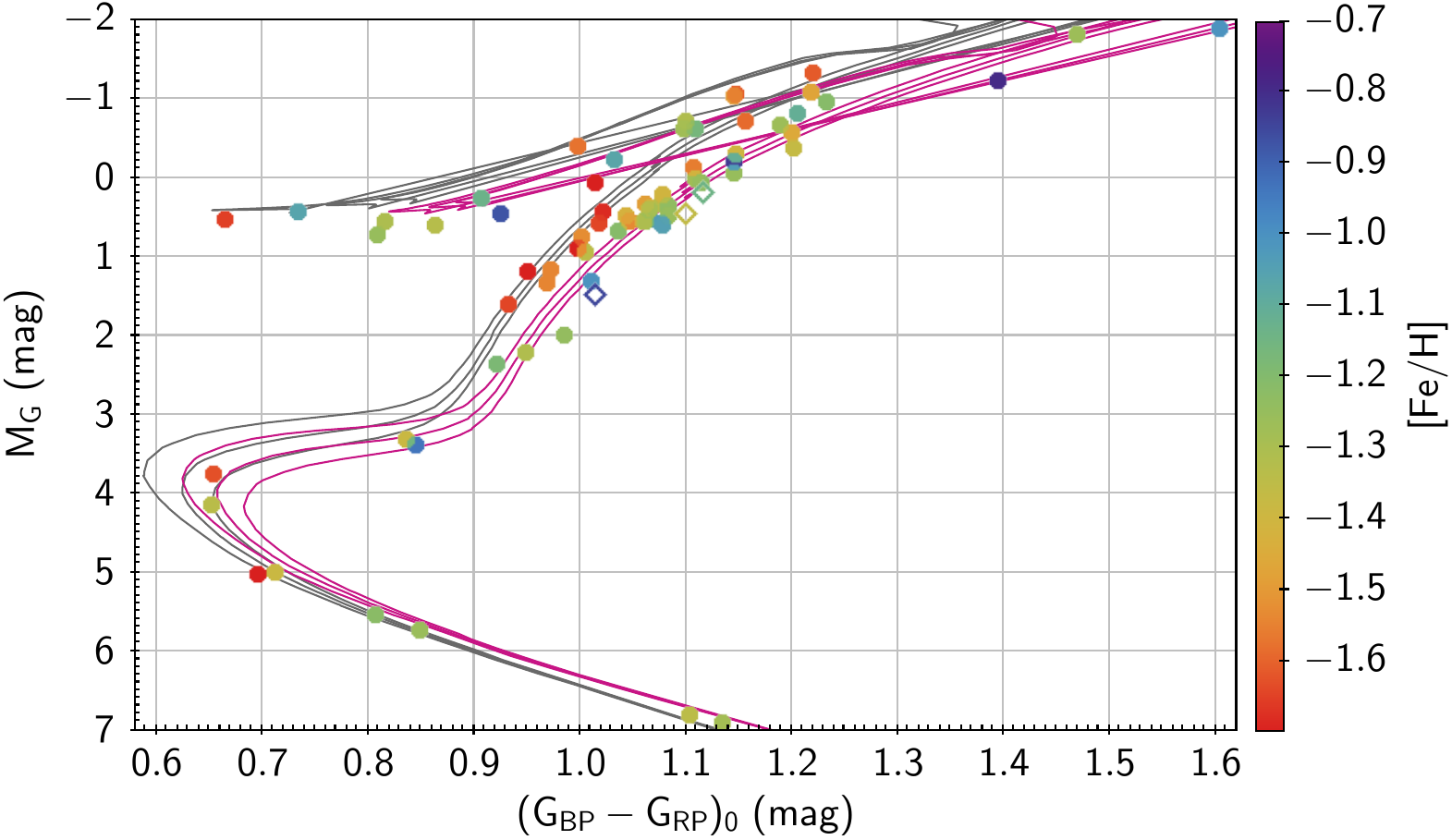}
   \caption{CMD diagram ${\rm{M_G}}$ vs.\ ${\rm (G_{BP}-G_{RP})_0}$ 
   	for the (accreted) Icarus stars as shown in Figure~\ref{fig:fig3}. 
    We show $71$~Icarus stars with high quality photometry:  68~objects have been corrected for extinction 
    using 3D~maps from \citet[][]{Vergely+2022} (filled circles) while for 3~objects (open diamonds) more accurate $A_V$ estimats by means of the EXOFASTv2 code \citep[][]{Eastman2019} are derived.
    These stars are color coded according to metallicity [Fe/H]. 
    Isochrones of ages $10, 12, 14~\rm{Gyr}$ for $\rm{[M/H]}=-1.0$ (magenta), and $\rm{[M/H]}=-1.3$ (gray) are from \citet[][]{Bressan2012} and \citet[][]{Marigo2017}. 
		}
   \label{fig:fig6}
 \end{figure*}

Figure~\ref{fig:fig6}  shows the color magnitude diagram (CMD) 
${\rm{M_G}}$ vs.\ ${\rm (G_{BP}-G_{RP})_0}$ of 71~Icarus members having accurate multi-band photometry (G, G$_{\rm BP}$, G$_{\rm RP}$) according to the criterion based on the corrected BP/RP flux excess factor $|C^*| < \sigma_{C*}$  defined by \citet[][]{Riello}. 68~stars have been de-reddened using the \citet[][]{Vergely+2022} extinction map\footnote{Available at \href{https://explore-platform.eu/sdas/about/gtomo}{G-TOMO}} with a resolution of 10~pc and a volume of 3~kpc $\times$ 3~kpc $\times$ 0.8~kpc. 

For three stars\footnote{{\it Gaia} Source Ids: 1119108395316741376, 6048584786354506752, and 5381152537909118976.} for which the 3D~map predicts an unrealistic low reddening, we estimated a more accurate extinction $A_V$ by means of the EXOFASTv2 code \citep[][]{Eastman2019} that performs a photometric fit of the stellar spectral energy distribution (SED). Our global fit is based on synthetic SDSS magnitudes (ugriz) derived from the {\it Gaia} BP/RP spectra \citep[GSPC;][]{GaiaCollab.2023:Montegriffo} combined with near-IR photometry (J, H, K$_s$) from 2MASS \citep[]{Cutri2003} and (W1, W2) magnitudes from AllWISE and CatWISE \citep[][]{Cutri2014, Eisenhardt2020}. In addition, the APOGEE astrophysical parameters ($T_{\rm eff}$, $\log g$, [Fe/H]) and the {\it Gaia} corrected parallax $(\varpi - ZP)$ were used as additional priors to solve the temperature-extinction degeneracy of the three sources. 

Finally, the extinction values $A_V$ provided by the map and by the EXOFASTv2 code have been converted to the filters of interest using the coefficients $A_{G}/A_{V}=0.836$, $A_{BP}/A_{V}=1.083$ and $A_{RP}/A_V=0.634$ \citep[][]{Mazzi} computed for a G2V star using the \citet[][]{ODonnel} mean interstellar extinction curve with $R_V = 3.1$. 

\subsection{Isochrone fitting}\label{sec:5.1}
Stars in different stellar phases provide different constraints about age and metallicity. The sample is dominated by RGB stars, with minor components of main sequence (MS), sub-gaint branch (SGB stars), horizontal branch (HB) and asymptotic giant branch (AGB) stars. To guide the eye and explore the age and metallicity of Icarus' populations, we over-plot the PARSEC-COLIBRI \citep[][]{Bressan2012, Marigo2017} solar-scaled ($\rm{[\alpha/Fe]=0.0}$) isochrones\footnote{Trasmission curves are from \citet[][]{MAWeiler2018}.} with stellar age $10$, $12$, and $14~\rm{Gyr}$ for metallicities  $\rm{[M/H]}=-1.0$ (magenta) and $\rm{[M/H]}=-1.3$ (gray).

Overall the proposed isochrones brackets well the color-magnitude position of:

\begin{itemize}
    \item the 6 sub-dwarfs (${\rm{M_G}>5}$);
    \item the 2 MS stars located near the Turn-Off (TO) region (${\rm{M_G}\approx 4}$);
    \item the 2 SGB stars located at ${\rm{M_G}\approx 3.5}$ and ($G_{BP}-G_{RP})_0  \approx 0.85$. 
\end{itemize}

The sub-dwarfs location is mainly sensitive to the metallicity. The isochrones with $\rm{[M/H]}=-1.3$ encompass all sub-dwarfs except one (which is indeed the most metal-poor of the six), corroborating the average spectroscopic metallicity of the sub-dwarf sample ($\rm{[Fe/H]}\approx -1.41$). On the other hand, TO and SGB regions are sensitive to both metal abundance and age. Guided by isochrones, it is evident that the 2 TO stars are old, with the precise age depending on their metallicity. From this point of view, spectroscopic measurements for the two TO stars suggest $\rm{[Fe/H]}=-1.33$ and $\rm{[Fe/H]}=-1.62$. Adopting the isochrones with metallicity $\rm{[M/H]}=-1.3$ for the former TO star, we infer a minimum age of 12~Gyr. Since the latter TO star is more metal poor, its age is probably even older. Concerning the 2~SGB stars, the measured metallicities are $\rm{[Fe/H]}=-0.93$ and $\rm{[Fe/H]}=-1.38$. Adopting the isochrones with metallicity $\rm{[M/H]}=-1.0$ for the former SGB star, we infer again a minimum age of 12~Gyr. Adopting the isochrones with metallicity $\rm{[M/H]}=-1.3$ for the latter SGB star, we infer a nominal age of 14~Gyr. 

Overall, the isochrones encompass well also the bulk of the RGB stars. While the inferred ages are old and roughly consistent with the previous determinations, the predicted $\rm{[M/H]}$ are higher than the spectroscopic $\rm{[Fe/H]}$ by about 0.3~dex. Indeed, the average $\rm{[Fe/H]}$ of the RGB stars near the isochrones with $\rm{[M/H]}=-1.0$ is $-1.28$, while the average $\rm{[Fe/H]}$ of the stars near the isochrones with $\rm{[M/H]}=-1.3$ is $-1.57$. However, Icarus's stars are generally alpha enhanced with an average $\rm{[\alpha/Fe]}$ (RGB stars only) of about $+0.27$, whereas the adopted stellar models are solar-scaled. \citet[][]{Salaris} have shown that $\alpha$-enhanced stellar isochrones can be well mimicked in CMDs by solar-scaled ones with the same total metallicity $\rm{[M/H]}$. Using the relation from \citet[][]{Salaris} and plugging $\rm{[\alpha/Fe]}=+0.27$, we find $\rm{[M/H]}-\rm{[Fe/H]}\approx +0.2$ dex, which explains most of the discrepancy.

\section{Comparison to Simulations}\label{sec:6}

\begin{figure*}
	\centering
	\gridline{\fig{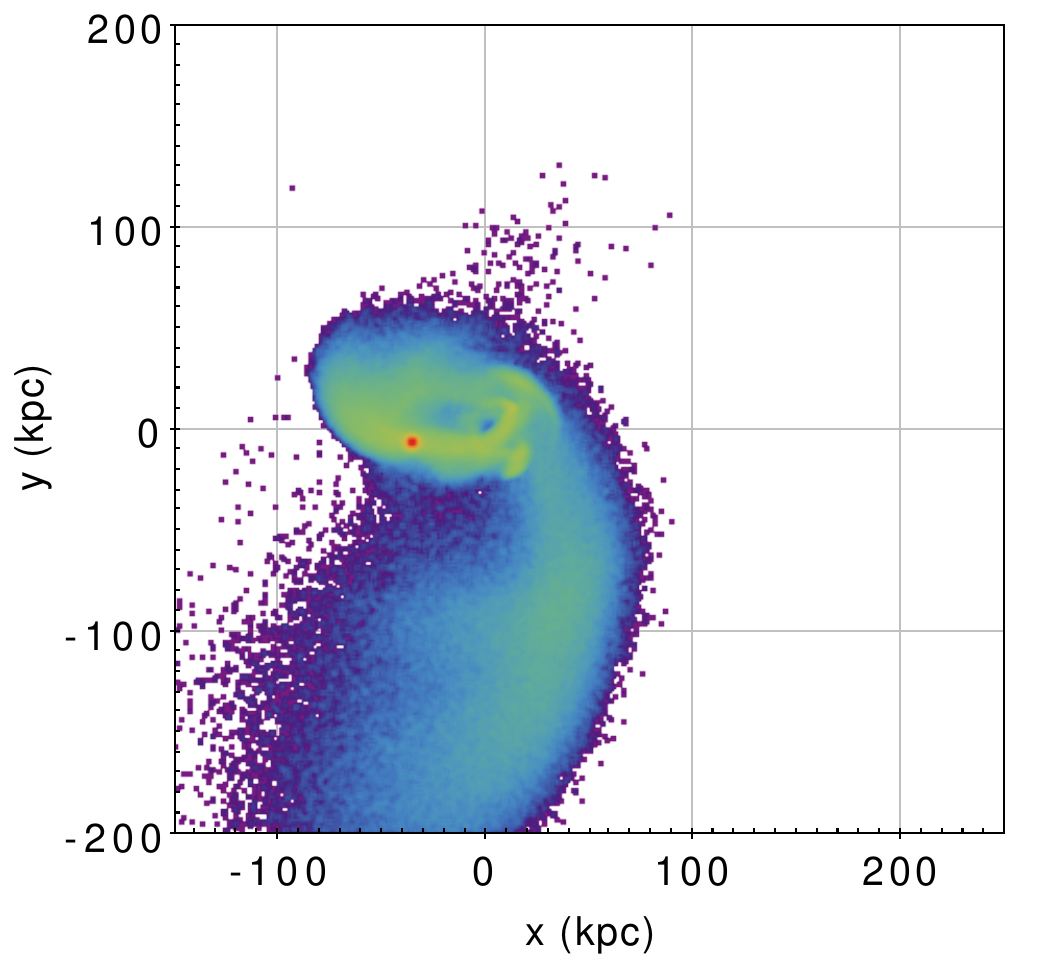}{0.33\textwidth}{}
              \fig{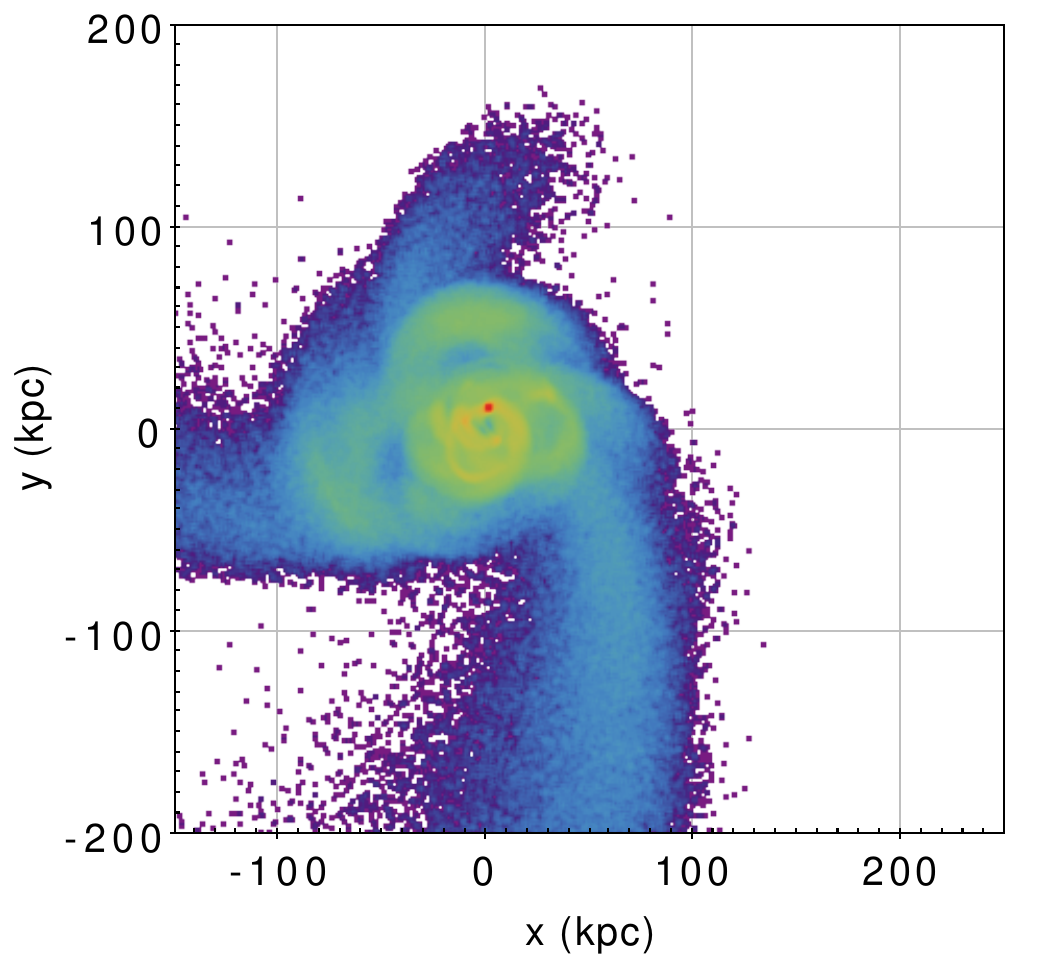}{0.33\textwidth}{}
              \fig{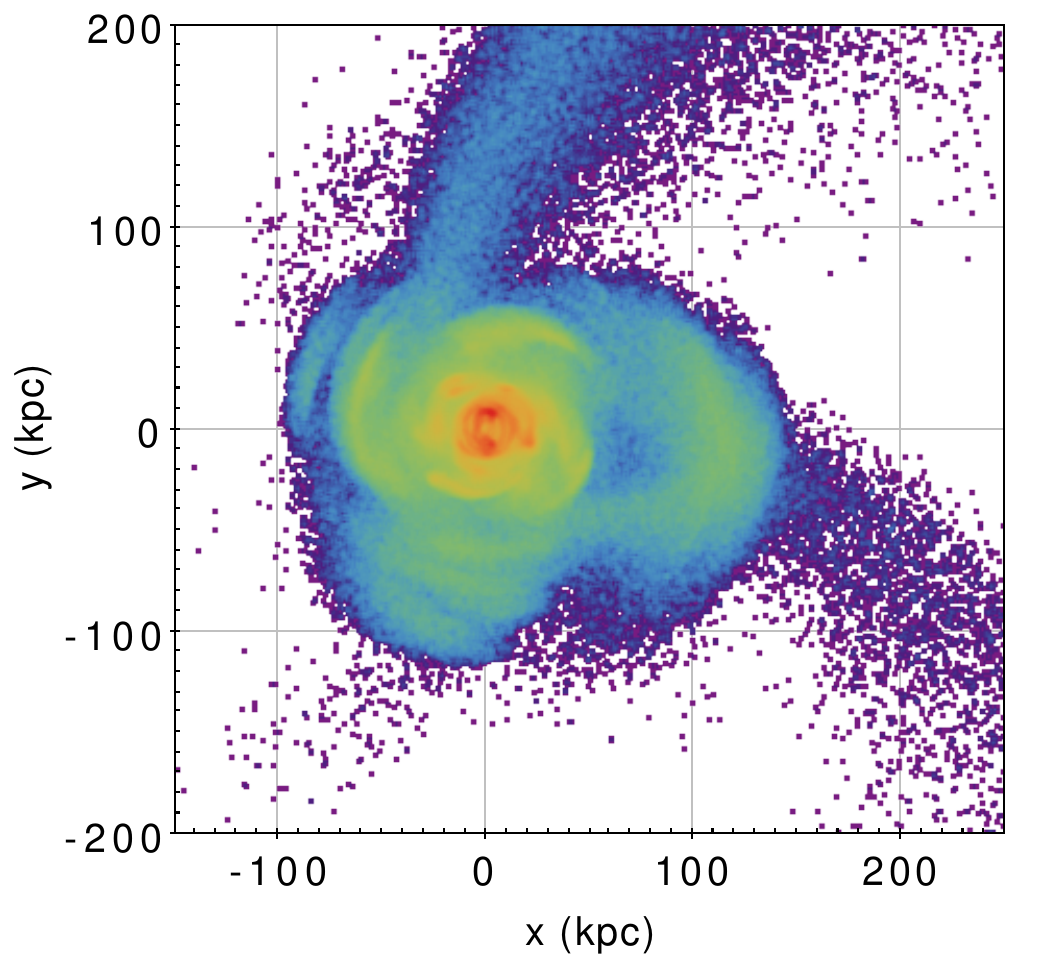}{0.33\textwidth}{}
              }
	\gridline{\fig{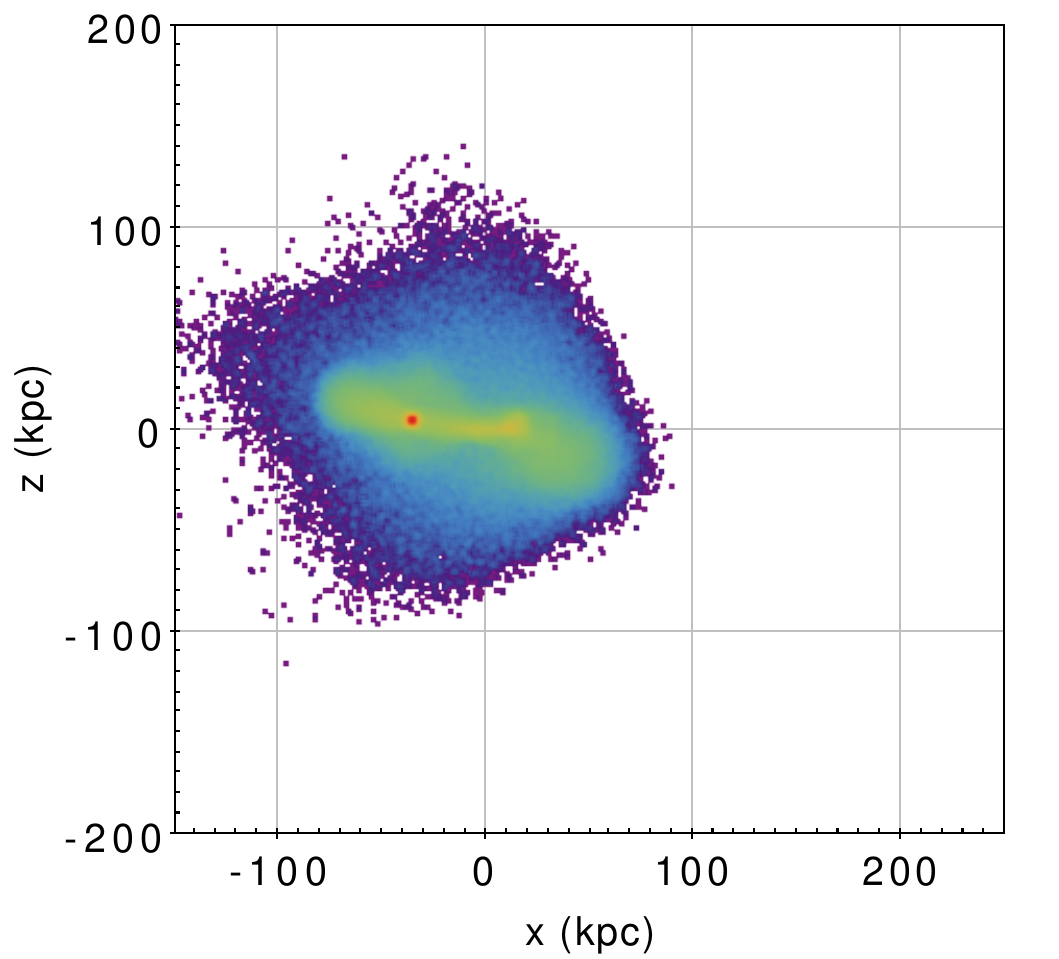}{0.33\textwidth}{}
              \fig{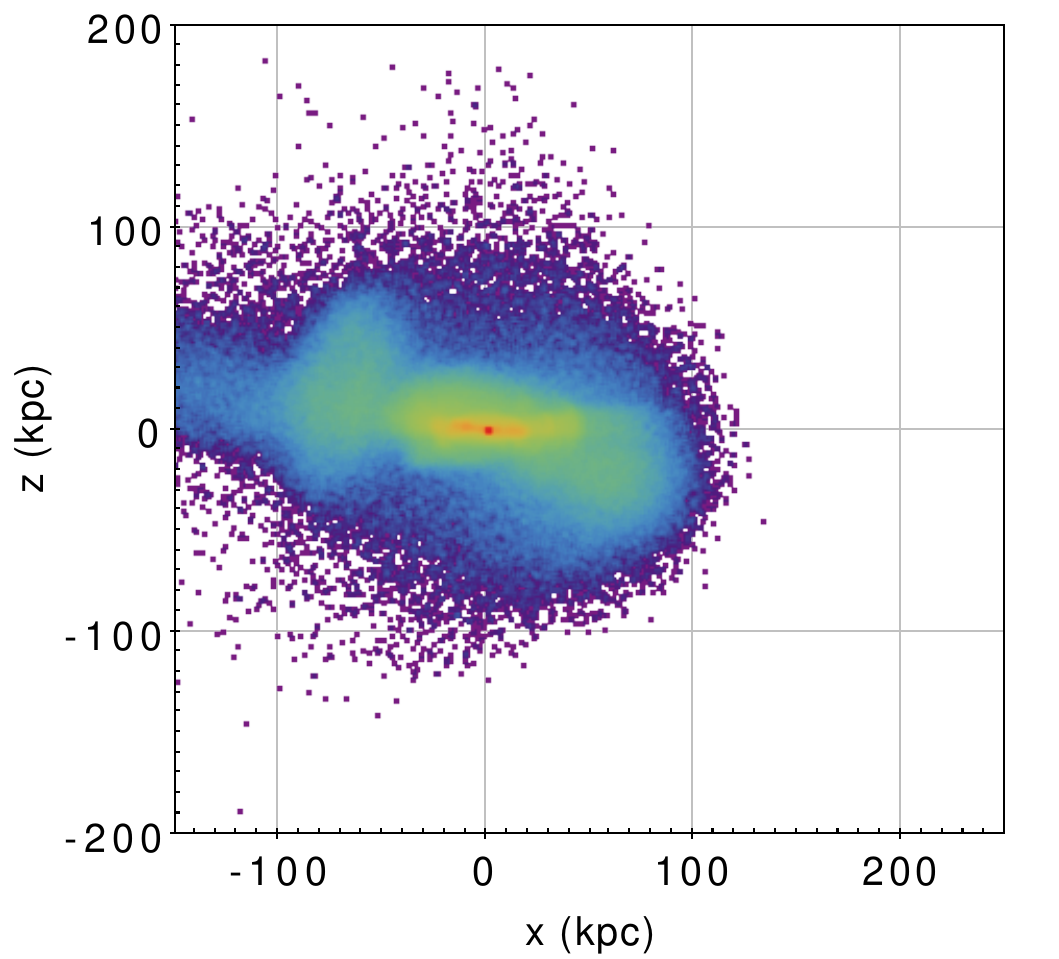}{0.33\textwidth}{}
              \fig{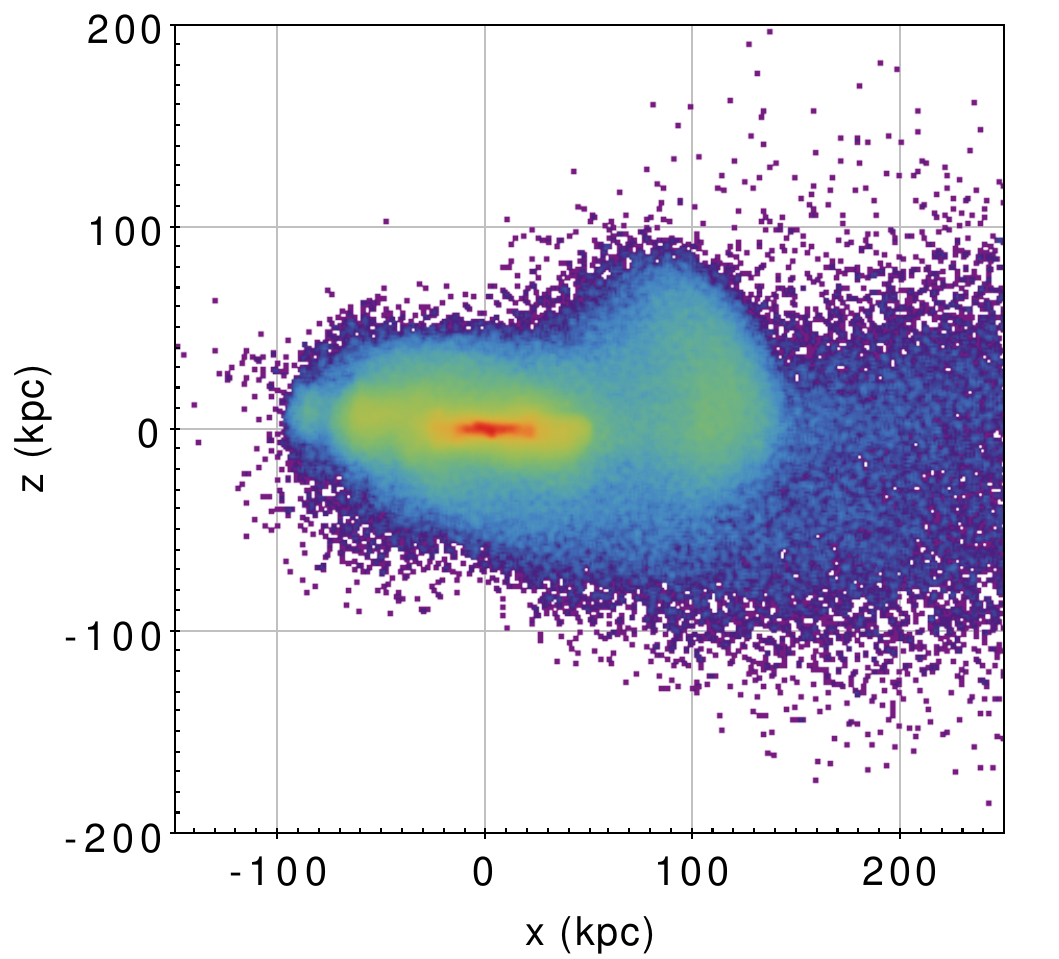}{0.33\textwidth}{}
              }
	\caption{N-body numerical simulation by \citet[][]{Murante2010} of the disruption of a LMC-like galaxy, on a 10$^\circ$-inclination prograde orbit, progressively destroyed by the gravitational pull of a MW-like galaxy 
	\citep[see also][]{Murante2010,ReFiorentin2015, ReFiorentin2021}. 
    Top panels: the spatial distribution in the x-y plane are shown, from left to right, after T=1.8~Gyr, T=3~Gyr, and the final configuration for the particles of the $10^\circ$-inclination satellite. 
	Bottom panels: as above, but for the x-z plane.
          }
	\label{fig:fig7} 
\end{figure*} 

In order to validate the {\it ex situ} origin of Icarus, 
we compare its newly defined members to 
the high-resolution N-body numerical simulation of a minor merger 
between a MW-like galaxy (``primary" system)
and a $10^\circ$-inclination prograde satellite published by 
\citet[][]{Murante2010} and analyzed by \citet[][]{ReFiorentin2015, ReFiorentin2021}.
Having stellar mass $M^*_{\rm satellite} \sim 10^9 M_\sun$ and 
virial mass ratio $M_{\rm primary}/M_{\rm satellite} \sim 40$, 
this simulation is representative of the interaction and accretion with a satellite similar to the Large Magellanic Cloud (LMC).

The system is left to evolve for 4.63~Gyr (about 16~dynamical timescales of the main halo) when the satellite has completed its merging with the primary halo. 
Figure~\ref{fig:fig7} shows the marginal distributions, $x$-$y$ and $x$-$z$, of the snapshots at T=1.8~Gyr, T=3~Gyr, and the final configuration, which evidence the spatial evolution of the satellite stellar particles.

Due to the strong dynamical friction affecting the low inclination prograde satellite, such system quickly loses its orbital energy and proceeds to the inner regions of the main halo towards the Galactic plane. 
During multiple passages, stellar debris are released with disk-like kinematics. 

 \begin{figure*}
	\centering
	\gridline{\fig{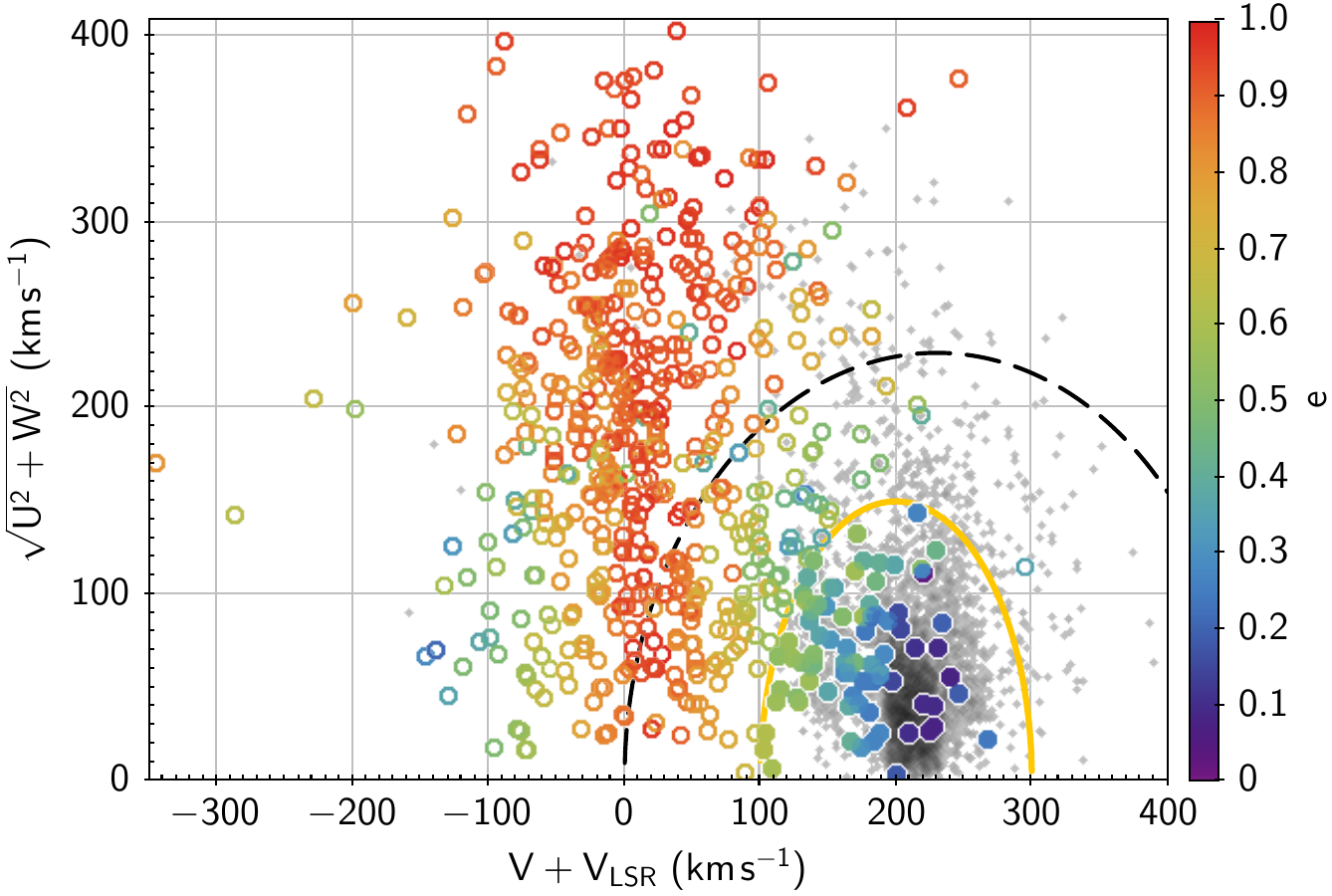}{0.5\textwidth}{}
              \fig{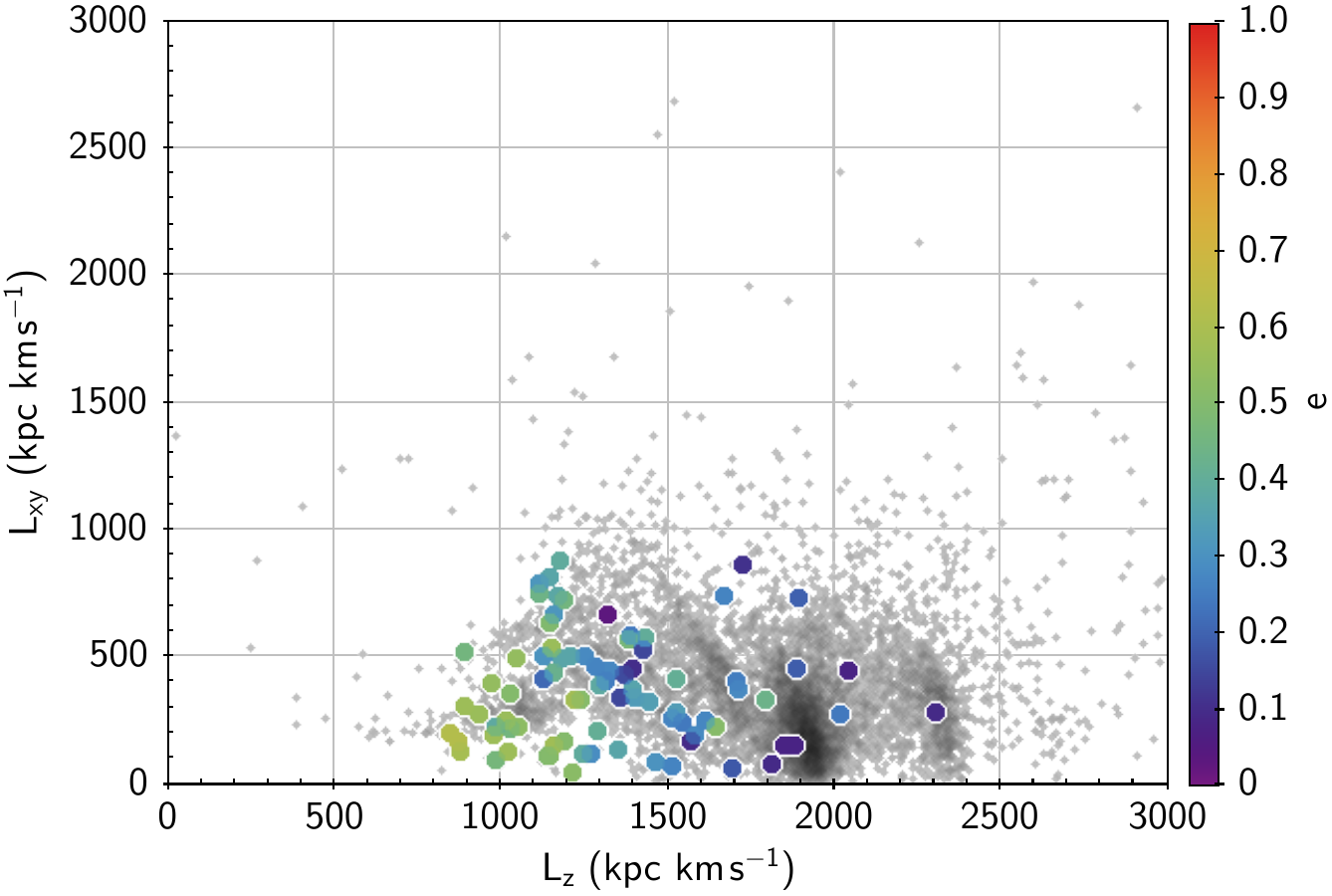}{0.5\textwidth}{}
              }
	\caption{Comparison to Simulations. 
	Left panel: Toomre diagram of the chemically selected accreted/unevolved sample, color coded accorging to eccentricity, as in Figure~\ref{fig:fig3}. 
   	Along Icarus stars (filled dots), GSE, Thamnos, and HS members are present. 
  	The (conservative) kinematical selection threshold for halo stars, 
    $\vert\vert{\bf v}-{\bf v}_{\rm LSR}\vert\vert> 230$~km~s$^{-1}$, 
	is represented by the dashed line.   
	The locus of Icarus is shown with solid line. 
	The debris of the simulated $10^\circ$-inclination prograde satellite produced by \citet[][]{Murante2010} and analyzed in \citet[][]{ReFiorentin2015, ReFiorentin2021}
	are overplotted (diamonds). 
	Right panel: space of adiabatic invariants, $L_{xy}$-$L_{z}$, for Icarus stars. 
	Again, the debris of the simulated $10^\circ$-inclination prograde satellite are overplotted for comparison (diamonds).
          }
	\label{fig:fig8}
\end{figure*} 

In the following, we compare our results 
with the kinematical signature left by this simulation after selecting particles in a sphere of 3~kpc radius centered at the ``Sun" ($x$ = 8.2~kpc from the ``Galaxy" center). 
The left panel of Figure~\ref{fig:fig8} shows the Toomre diagram 
for the full sample of $622$~metal-poor accreted/unevolved stars color coded by eccentricity; 
filled circles evidence Icarus members as in Figure~\ref{fig:fig3}.  
Above the dashed line is the region usually associated with the Galactic halo, $\vert\vert{\bf v}-{\bf v}_{\rm LSR}\vert\vert> 230~\rm{km~s}^{-1}$ which is dominated by GSE; 
the solid line represents the locus of Icarus defined by Eq.~\ref{eq:6}. 
The simulated stars from the 
$10^\circ$-inclination prograde satellite 
selected in the ``solar neighborhood" ($d<3~\rm{kpc}$) are the grey diamonds. 
The bulk of the debris of the satellite are found in the Icarus locus. 
This result further supports the accreted origin of Icarus and helps to characterize its progenitor.

We also explore the space of ``integrals of motion'' defined by the components of angular momentum in and out the Galaxy's disk. 
The right panel of Figure~\ref{fig:fig8} shows the plane $L_{xy}$-$L_{z}$ for Icarus members and for the same debris of the accreted satellite shown in the Toomre diagram. 
We remark the presence of several kinematical substructures among the satellite particles similar to those observed in the distribution of the Icarus stars. 
In addition, we point out that a few other prograde kinematical groups detected by \citet[][see e.g.\ Figure~4]{ReFiorentin2021} are also associated to the dynamical substructures left by the Icarus progenitor.


\section{Discussion and Conclusions}\label{sec:7}

Detection and characterization of an accreted dwarf galaxy in the Galactic disk is a challenging task, to which {\it Gaia} in continuous synergy with ground-based wide field and high-resolution spectroscopic surveys play a crucial role. 

The discovery of five substructures with disk kinematics was reported by \citet[][]{ReFiorentin2021} that analysed a sample of 163~low-metallicity 
stars with relative velocity less than 15~km~s$^{-1}$ within 2.5~kpc of the Sun. These include the Icarus stream (Group~4), the fast-rotating stellar stream that  
\citet[][]{ReFiorentin2021} associated to the debris of a dwarf galaxy by means of a high resolution N-body numerical simulation of an accreted satellite with a stellar mass of $\sim 10^9 M_\sun$ on an initial prograde low-inclination orbit, $10^\circ$.
In addition, they argued that the other four substructures (Group~3, 5, 6, 7) were either streams previously released by the same progenitor of Icarus, or remnants from different satellites accreted along an initial prograde orbit, but with higher inclinations.

In order to gain more insight into the origin of Icarus as remnant of a disrupted dwarf galaxy rather than a signature of the Milky Way’s disk formation and evolution, we have complemented {\it Gaia}~DR3 with high resolution spectroscopic data from APOGEE~DR17 and GALAH~DR3.
We have extracted a sample of 622~stars within 3~kpc of the Sun
in the chemical space of the accreted/unevolved stars based on their 
[Mg/Mn] vs.\ [Al/Fe] and [Mg/Fe] vs.\ [Fe/H] distributions 
and explored their chemo-dynamical properties. 

In Sect.~\ref{sec:3}, by means of Eq.~\ref{eq:5} and Table~\ref{table:1}, 
we have classified 376~GSE sources and 
81~newly identified members\footnote{The list of Icarus' members is available on request from the corresponding author, P. Re~Fiorentin.} of Icarus, characterized by peculiar high and low orbital eccentricity, respectively. 

In Sect.~\ref{sec:4} we further investigated the chemical properties of Icarus and GSE stars in order to validate their extragalactic origin. 
In particular, we explored the chemical distributions 
[Ni/Fe] vs.\ [(C+N)/O], 
[Y/Eu] vs.\ [Fe/H], [Eu/Mg] vs.\ [Fe/H], and [Ba/Y] vs.\ [Fe/H], 
where accreted debris and {\it in situ} populations can be disentangled \citep[e.g.,][]{Montalban, Carrillo}.
Indeed, in these diagnostic chemical spaces, 
the GSE and Icarus candidates appear well separated from the bulk of MW stars, as well as they show a significant overlap that supports the Icarus {\it ex situ} origin. 

We also explored the distribution 
[Ba/Mg] vs.\ [Mg/H] that evidences the contribution from AGB stars to the chemical enrichment (see Figure~\ref{fig:fig5}, bottom right panel). Indeed, Icarus stars seem to show an increasing trend similar to that observed in other dwarf systems, 
such as Fornax \citep[][]{Letarte}, Ursa Minor \citep[][]{Sestito2023}, and Sagittarius \citep[][]{Sestito2024b}.

As described in Sect.~\ref{sec:5}, the comparison between the CMD of Icarus members and PARSEC-COLIBRI isochrones reveals that these stars are 
older than 12~Gyr. This age estimate sets an important upper limit on the epoch of accretion: assuming quenching of star formation within the Icarus progenitor at the time of merging the CMD of the stream is a snapshot of that event.

In Sect.~\ref{sec:6}, we compare the kinematical properties of the Icarus members to a high resolution N-body simulation representative of the merger between an LMC-like satellite and a MW-like galaxy \citep[][]{Murante2010}. We found a remarkable consistency in the Toomre diagram and angular momentum distribution $L_{xy}$ vs.\ $L_{z}$ between the Icarus stream and the stellar debris, selected in a local volume ($d<3$ kpc of the ``Sun"), stripped from a satellite on an initial $10^\circ$-inclination prograde orbit. \\

This result is also consistent with the recent high resolution N-body simulations analyzed by \citet[][]{Mori}; in fact, by inspection of the left panels of their Figs.\ 5-6, 
we point out the presence of metal-poor stars with disk kinematics due to debris of a satellite accreted on a prograde and zero-inclination orbit ($\Phi=0$).

The accretion scenario is also supported by high resolution cosmological simulations of MW-like galaxies 
\citep[][]{Sestito2021,Santistevan,Carollo, Pinna}.
In particular, \citet[][]{Santistevan} 
found that the significant excess of metal-poor stars with 
prograde vs.\ retrograde orbits
derives from debris released by 
a single primary merger of Small/Large Magellanic Cloud (SMC/LMC)-mass satellite.
Such primary merger is the most important contributor of {\it ex situ} with respect to the {\it in situ} stars with $-1.7<\rm{[Fe/H]}<-0.7$ that corresponds exactly to the metallicity range of our input catalog.

Therefore, even though we cannot exclude the contamination of unevolved low metallicity {\it in situ} stars, we argue that the Icarus stream is consistent with debris of a low-inclination prograde dwarf galaxy with a stellar mass $\sim 10^9 M_\sun$. In any case, we remark that this scenario assumes the existence of a protogalactic disk formed {\it in situ} at ages $>12$~Gyr \citep[cf.\ ][]{XiangRix2022}. A similar scenario was suggested by \citet[][]{Carter} that proposed the accretion of low-$\alpha$ stars from a co-rotating dwarf galaxy onto a primordial high-$\alpha$ disk.

However, the origin of metal-poor stars with disk kinematics is currently a matter of lively debate in the astronomical community. 
We think that such controversial interpretations could derive from the different selection criteria that may generate kinematically or chemically biased samples.

For instance \citet[][]{Mardini} selected a subset of 7\,127 MWTD stars with 
[Fe/H]$<-0.8$, $140 < V_\phi < 160$~kpc~km~s$^{-1}$, $ 0.3<e<0.6$, $z_{\rm max} < 3$, and $  J_\phi/J_{\rm tot} < -0.98$. 
They proposed an accreted origin for these objects that they dubbed the Atari disk. 
Actually, we notice that their kinematically selected sample is basically included in our Icarus stream as shown in Figure~\ref{fig:fig3}. In fact, 8 of the 81 Icarus members satisfy also the criterions given by \citet[][]{Mardini}.

Recently, \citet[][]{Malhan2024} identified two overdensities in the distribution $(E,L_z)$ of prograde metal-poor stars with [Fe/H]$<-1$. The origin of these structures, 
named ``Shakti" and ``Shiva", is still uncertain as their chemical properties do not clearly disentangle between dynamical signatures of {\it in situ} stars due to the MW bar and remnants of ancient Galactic building blocks that merged 
12-13~billion years ago. Indeed, we remark that 
the chemo-dynamical properties of Shakti  overlap with Icarus\footnote{From our 35/81 Icarus members with Shakti kinematics 
($L_{xy}<600$~kpc~km~s$^{-1}$ and 
$800<L_{z}< 1300$~kpc~km~s$^{-1}$) 
we estimate mean values of   
$L_{z}= 1100$~kpc~km~s$^{-1}$, 
[Mg/Fe] $= +0.26$, 
eccentricity $e= 0.48$, 
$z_{\rm max} = 1.5\pm 0.8$~kpc.
Such values match pretty well median values of 
$L_{z}= 1000$~kpc~km~s$^{-1}$ 
(Malhan private communication), 
[Mg/Fe] $= +0.3$, 
$e= 0.4$, 
$z_{\rm max} = 3.2$~kpc 
given for Shakti in Table~1 by \citet[][]{Malhan2024}.} and correspond to the substructure dubbed ``Group 5'' by \citet[][]{ReFiorentin2021}.

In addition, other studies claim that these 
metal-poor stars with disk kinematics were formed {\it in situ} and represent the early stellar disk of the MW. 

For instance, \citet[][]{Feltzing} 
select a prograde sample ``Ib" with $L_z/J_{\rm tot}>0.6$ and chemical abundances [Mg/Mn] vs.\ [Al/Fe] typical of an accreted origin; these stars are characterized by a mean eccentricity of 0.37 with a standard deviation 0.16, very similar to the values that we have estimated for Icarus (see Table~\ref{table:2}). This is not surprising because their sample of 379 objects include 37~Icarus members (Feuillet, private communication). 

\citet[][]{Feltzing} associated the stars of group ``Ib" to a remnant of the primordial Galactic disk formed from a chemically unevolved medium before the formation of the canonical thick disk, because of the low [Al/Fe] but higher [Mg/Fe] abundance of these stars with respect to those with halo kinematics.
However, the small difference between $\langle$[Mg/Fe]$\rangle\simeq +0.27$ and $+0.21$ that we estimate respectively for Icarus and GSE (see Table~\ref{table:3}) could also be explained by a dominant accreted population with a contamination of {\it in situ} stars.

\citet[][]{Fernandez-Alvar} focus on metal-poor stars with $-2<$[Fe/H]$<-0.7$ and thin disk-like kinematics: $L_z/J_{\rm tot}>0.95$, $(J_z-J_r)/J_{\rm tot}<0.05$, and $|z_{\rm max}|<1.5$~kpc. 
They found that the bulk of these objects are confined within $|z_{\rm max}|<0.75$~kpc and concluded that this population represents the metal-poor edge of the ``thin" disk.
In particular, 
the distribution [Mg/Mn] vs.\ [Al/Fe] of the subset with [Fe/H]$<-1$ shows that these stars, as Icarus\footnote{We also point out the significant overlap, as 37 of the 81 Icarus members satisfy the kinematical selection criterions of the \citet[][]{Fernandez-Alvar} sample.}, have been formed in a less chemically evolved medium than the canonical thick disk.

Similarly, \citet[][]{Nepal} analyzed a sample of metal-poor stars in thin disk-like orbits:  
[Fe/H]$<-1$, 
$V_\phi > 180$~km~s$^{-1}$, and $|z_{\rm max}|<1$~kpc.
For this kinematically selected sample, the authors estimated 
$\langle V_\phi\rangle\simeq 218\,\rm{km~s}^{-1}$, $\sigma_{Vz}\simeq 20-25~\rm{km~s}^{-1}$ and 
\verb+StarHorse+ ages up to 13~Gyr. 
Then, \citet[][]{Nepal} concluded that this ancient population represents the remnant of the primordial thin disk that would have been formed in an inside out manner less than 1~Gyr from Big Bang.
According to these authors, such scenario is supported by the Toomre diagrams shown in the top panels of their Figure~2, where old stars with 10~Gyr $<$ age $<$ 14~Gyr and disk-like orbits have been found at all metallicities, $-2.5<$[Fe/H]$<+0.5$. 

However, we propose a different interpretation that can explain these results:
the old metal-poor stars could be the debris of a dwarf galaxy that merged with stars belonging to the primordial MW disk.
This scenario is supported by the fact that 10 of our 81~Icarus members, corresponding to the high velocity tail, satisfy the selection criterions given by \citet[][]{Nepal}.

This hypothesis is also consistent with the distribution of [Mg/Mn] vs.\ [Al/Fe] shown in Fig. 7 by \citet[][]{Fernandez-Alvar}, 
where thin disk-like stars with [Fe/H]$<-1$ 
populate the upper left region associated to accreted or chemically unevolved stars. Conversely, the bulk of stars with 
$-1<$[Fe/H]$<+0.5$ populate the region associated to {\it in situ}  thin and thick disk.

On the other hand, recent studies still support a dominant {\it in situ} formation of the ancient metal-poor MW disk \citep[][]{Bellazzini,Conroy}. 
In particular, \citet[][]{Conroy} 
argued that the MW disk has been formed {\it in situ} through various phases with different star formation efficiency and that the contamination of accreted stars with prograde orbits is negligible. 
This scenario is compatible with the ``dual origin" of the MW thick disk proposed by \citet[][]{Grand} on the basis of mergers similar to GSE on MW-like galaxies from  AURIGA simulations. 
These simulations show that thick disks can be formed from {\it in situ} starbursts triggered by the accreted gas, as well as by dynamical heating of a pre-existing thin disks. In addition, a fraction of proto-disk stars are also scattered into the halo (the ``Splash").

Finally, other studies claim that metal-poor stars with disk orbits may be the tail of a prograde component of the halo, possibly originated by bar resonances or accreted building blocks \citep[][]{BelokurovKravtsov, Dillamore, AnkeAA, Zhang}. 

The prograde halo hypothesis has been also considered by \citet[][]{Gonzalez} that analyzed 36\,000 stars with [Fe/H]$<-1.7$ and disk kinematics from the Pristine survey. These authors suggest that this population may include stars belonging to the metal-weak thick disk, as well as to the prograde halo. However, they point out the need of high resolution spectra to clarify the {\it in situ}/{\it ex situ} origin of this population.

In this context, 
we complemented dynamical properties and age estimates from {\it Gaia}~DR3 with a detailed chemical analysis of the APOGEE and GALAH elemental abundances 
[Fe/H], [Mg/Fe], [Mn/Fe], [Al/Fe], 
[(C+N)/O] vs.\ [Ni/Fe], 
[Y/Eu], [Eu/Mg], [Ba/Y], and [Ba/Mg] to better identify and characterize the Icarus stream in comparison to GSE. 
This gave us more insight into the origin of Icarus as a remnant of a disrupted dwarf galaxy thanks to both the more detailed chemical abundances of its members just recalled, and through a reanalysis of our tailored N-body  results, as well as other suitable dynamical simulations published in the recent literature \citep[][]{Mori, Sestito2024}. 
We found that Icarus and GSE show chemo-dynamical properties typical of two accreted low-mass systems with different initial orbital parameters and diverse chemical histories. 

Yet, at this point, we cannot exclude the possibility of a primordial disk origin. 
Accurate stellar ages, improved astrometry (from {\it Gaia}'s next releases), increased spectra (from, e.g., APOGEE, GALAH, WEAVE \citep[][]{WEAVE}, 4MOST \citep[][]{4MOST} surveys), and dedicated spectroscopic follow-up whenever necessary, will confirm new members, thus strengthening the {\it ex situ} origin of Icarus as the first prograde satellite galaxy in the disk of the Milky Way.

\acknowledgments 
We are grateful to the anonymous referee for a careful reading of this manuscript and for the useful comments that helped us improve the original manuscript. 
We wish to thank W.\ Beordo, C.\ Chiappini, A.\ Curir, D.\ Feuillet, D.\ Horta, G.\ Kordopatis, A.\ Queiroz, 
I.\ Gonz\'alez Rivera de La Vernhe, and E.\ Fern\'andez-Alvar, for helpful discussions. 

P.~R.F., A.~S., and M.~G.~L.\ are supported in part by the Italian Space Agency (ASI) through contract 2018-24-HH.0 and its addendum 2018-24-HH.1-2022 to the National Institute for Astrophysics (INAF).
M.~C. acknowledges the support of INFN ``Iniziativa specifica TAsP". 
S.~V. thanks ANID (Beca Doctorado Nacional, folio 21220489) and the Millennium Nucleus ERIS (ERIS NCN2021017) for the funding. 

This work has made use of data from the European Space Agency (ESA) mission {\it Gaia} (\href{https://www.cosmos.esa.int/gaia}{https://www.cosmos.esa.int/gaia}), 
processed by the {\it Gaia} Data Processing and Analysis Consortium (DPAC, \href{https://www.cosmos.esa.int/web/gaia/dpac/consortium}{https://www.cosmos.esa.int/web/gaia/dpac/consortium}). Funding for the DPAC has been provided by national institutions, in particular the institutions participating in the {\it Gaia} Multilateral Agreement. 

Funding for the Sloan Digital Sky Survey IV has been provided by the Alfred P. Sloan Foundation, the U.S. Department of Energy Office of Science, and the Participating Institutions. 
SDSS-IV acknowledges support and resources from the Center for High Performance Computing  at the University of Utah. The SDSS website is \href{www.sdss4.org}{www.sdss4.org}.
SDSS-IV is managed by the Astrophysical Research Consortium for the Participating Institutions of the SDSS Collaboration including the Brazilian Participation Group, the Carnegie Institution for Science, Carnegie Mellon University, Center for Astrophysics | Harvard \& Smithsonian, the Chilean Participation Group, the French Participation Group, Instituto de Astrof\'isica de Canarias, The Johns Hopkins University, Kavli Institute for the Physics and Mathematics of the Universe (IPMU) / University of Tokyo, the Korean Participation Group, Lawrence Berkeley National Laboratory, Leibniz Institut f\"ur Astrophysik Potsdam (AIP), Max-Planck-Institut f\"ur Astronomie (MPIA Heidelberg), Max-Planck-Institut f\"ur Astrophysik (MPA Garching), Max-Planck-Institut f\"ur Extraterrestrische Physik (MPE), National Astronomical Observatories of China, New Mexico State University, New York University, University of Notre Dame, Observat\'ario Nacional / MCTI, The Ohio State University, Pennsylvania State University, Shanghai Astronomical Observatory, United Kingdom Participation Group, Universidad Nacional Aut\'onoma de M\'exico, University of Arizona, University of Colorado Boulder, University of Oxford, University of Portsmouth, University of Utah, University of Virginia, University of Washington, University of Wisconsin, Vanderbilt University, and Yale University. 

This work made use of the Third Data Release of the GALAH Survey (Buder et al. 2021). The GALAH Survey is based on data acquired through the Australian Astronomical Observatory, under programs A/2013B/13 (The GALAH pilot survey); A/ 2014A/25, A/2015A/19, A2017A/18 (The GALAH survey phase 1); A2018A/18 (Open clusters with HER- MES); A2019A/1 (Hierarchical star formation in Ori OB1); A2019A/ 15 (The GALAH survey phase 2); A/2015B/19, A/2016A/22, A/2016B/10, A/2017B/16, A/2018B/15 (The HERMES-TESS program); and A/2015A/3, A/2015B/1, A/2015B/19, A/2016A/22, A/2016B/12, A/2017A/14 (The HERMES K2-follow-up program). We acknowledge the traditional owners of the land on which the AAT stands, the Gamilaraay people, and pay our respects to elders past and present. This paper includes data that have been provided by AAO Data Central (datacentral.aao.gov.au). 

This work made extensive use of TOPCAT \citep[][]{Taylor}.


{}

\end{document}